\documentclass[twocolumn]{aastex61}
\usepackage{mathrsfs, mathtools}

\begin{document}

\title{Dust Density Distribution and Imaging Analysis of Different Ice Lines in Protoplanetary Disks}

\correspondingauthor{Paola~Pinilla, Hubble fellow}
\affiliation{Department of Astronomy/Steward Observatory, The University of Arizona, 933 North Cherry Avenue, Tucson, AZ 85721, USA}
\email{pinilla@email.arizona.edu}

\author{P.~Pinilla}
\affiliation{Department of Astronomy/Steward Observatory, The University of Arizona, 933 North Cherry Avenue, Tucson, AZ 85721, USA}
\author{A.~Pohl}
\affiliation{Max Planck Institute for Astronomy, K\"{o}nigstuhl 17, D-69117 Heidelberg, Germany.}
\affiliation{Heidelberg University, Institute of Theoretical Astrophysics, Albert-Ueberle-Str. 2, D-69120 Heidelberg, Germany}
\author{S.~M.~Stammler}
\affiliation{University Observatory, Faculty of Physics, Ludwig-Maximilians-Universit\"{a}t M\"{u}nchen, Scheinerstr. 1, D-81679 M\"{u}nich, Germany} 
\author{T. Birnstiel}
\affiliation{University Observatory, Faculty of Physics, Ludwig-Maximilians-Universit\"{a}t M\"{u}nchen, Scheinerstr. 1, D-81679 M\"{u}nich, Germany} 

\begin{abstract}
Recent high angular resolution observations of protoplanetary disks at different wavelengths have revealed several kinds of structures, including multiple bright and dark rings. Embedded planets are the most used explanation for such structures, but there are alternative models capable of shaping the dust in rings as it has been observed. We assume a disk around a Herbig star and investigate the effect that ice lines have on the dust evolution, following the growth, fragmentation, and dynamics of multiple dust size particles, covering from 1\,$\mu$m to 2\,m sized objects. We use simplified prescriptions of the fragmentation velocity threshold, which is assumed to change radially at the location of one, two, or three ice lines. We assume changes at the radial location of main volatiles, specifically H$_2$O, CO$_2$, and NH$_3$. Radiative transfer calculations are done using the resulting dust density distributions in order to compare with current multiwavelength observations. We find that the structures in the dust density profiles and radial intensities at different wavelengths strongly depend on the disk viscosity.  A clear gap of emission can be formed between ice lines and be surrounded by ring-like structures, in particular between the H$_2$O and CO$_2$ (or CO). The gaps are expected to be shallower and narrower at millimeter emission than at near-infrared, opposite to model predictions of particle trapping. In our models, the total  gas surface density is not expected to show strong variations, in contrast to other gap-forming scenarios such as embedded giant planets or radial variations of the disk viscosity.
\end{abstract}


%%%%%%%%%%
\section{Introduction}     \label{introduction}
%%%%%%%%%%

Recent high angular resolution observations of protoplanetary disks revealed several examples of multiple ring-like structures,  for instance, in disks around HL\,Tau \citep{alma2015}, HD\,100546 \citep{walsh2014}, HD\,97048 \citep{ginski2016, plas2016, walsh2016}, TW\,Hya \citep{andrews2016, boekel2017}, HD\,163296 \citep{isella2016}, HD\,169142 \citep{fedele2017}, and RX\,J1615.3-3255 \citep{deboer2016}. These objects have a large range of properties, including  different stellar types and ages (from very young, $\lesssim1\,$My, to very old, $\sim10\,$My). 

The current literature for explaining dust rings and gaps in protoplanetary disks is very rich and includes zonal flows from the magnetorotational instability  \citep[MRI; e.g.,][]{johansen2009, uribe2011, dittrich2013, simon2014}, spatial variations of the disk viscosity \citep[e.g.][]{kretke2007, regaly2012, flock2015, pinilla2016}, secular gravitational instability \citep[e.g.][]{youdin2011, takahashi2014}, instabilities originating from dust settling \citep{loren2015},  self-sustained recycling of inner dust rings \citep{husmann2016}, particle growth by condensation near ice lines \citep{saito2011, ros2013, stammler2017}, sintering of dust particles that inhibits dust growth near the ice lines \citep{okuzumi2016}, and planet-disk interaction \citep[e.g.][]{rice2006, zhu2011, gonzalez2012, pinilla2012, dipierro2016, dong2016,  rosotti2016}. Although the latter explanation is the most widely used to interpret current observations, it is not a unique possibility, and several of the listed processes can play an important role during the disk evolution. 

To differentiate between all these models for the origin of rings and gaps, it is crucial to predict the behavior of the gas, as well as dust particles of different size, and compare with current observations of disks that cover wavelengths from optical to the millimeter emission. In this paper, we further investigate the effect that different ice lines have on the dust evolution. We predict the distributions of small (micron-sized) particles versus the distribution of large (millimeter-sized) grains and give imaging diagnostics to consider or exclude this scenario as the cause of the seen structures.

In this work, we consider the growth and fragmentation of dust particles during the disk evolution. The growth from micron-sized particles to larger bodies occurs as a result of sticking collisions. The sticking efficiency between pairs of dust grains depends on the Van der Waals forces, which are attractive forces between permanent, induced, or fluctuating dipoles. \cite{hamaker1937}  calculated the resulting Van der Waals force between two spherical particles as a function of their diameters and the distance separating them, such that the total attractive force is given by $\sim -C_H/r^6$. The Hamaker constant ($C_H$) is the sum of the dispersion, polarizability, and  orientation coefficients of the pairs in the interaction. Dispersion forces are exhibited by non-polar molecules because of the fluctuating moments of the nucleus and electrons of the atoms or molecules. For nonpolar molecules, the higher the contribution of the dispersion force to the Van der Waals force, the lower is the magnitude of the net Van der Waals force. For instance, carbon dioxide (CO$_2$) is a nonpolar molecule, and therefore the only contribution to the Van der Waals force is the dispersion force \citep{french2007}. As a consequence, when the mantle of  two dust particles is composed mainly of CO$_2$ ice (or another molecule with very low dipole moment, e.g. CO or silicates), the net attractive force between the two grains is weaker compared to the net attraction of particles constituted by other polar molecules, such as water or ammonia \citep[the total contribution of the dispersion force to the Van der Waals force is 24\% and 57\% for water and ammonia, respectively; e.g.,][] {mohanty1976, israelachvili1992}.

To reassemble the dependence of the dipole moment of the constituents on the net attractive force between dust particles, we simply assume that the threshold of the dust collision velocity to cause fragmentation of particles (the fragmentation velocity, $v_{\rm{frag}}$)  accordingly changes radially at the location  of one, two, or three ice lines. Specifically, we assume changes at the radial location where H$_2$O, NH$_3$, and CO$_2$ are expected to freeze out. In these models, we assume that the mantle of dust grains is mainly composed of the volatile that freezes out at a given location.  We analyze how the changes of the fragmentation velocity affect the final dust distribution of particles with 1\,$\mu$m to millimeter sizes. In order to compare with current multiwavelength observations of protoplanetary disks, we include radiative transfer calculations and calculate the radial profile of synthetic images at different wavelengths.

The organization of this paper is as follows.  In Sect.~\ref{sect:models}, we explain the details and assumptions of our dust evolution and radiative transfer models. Section~\ref{sect:results} and Sect.~\ref{sect:images} present  the results of the dust density distribution and radial profiles of synthetic images from different models. Section~\ref{sect:discussion} and Sect.~\ref{sect:conclusions} provide a discussion and conclusions, respectively. \vspace{1cm}

\section{Models and Assumptions} \label{sect:models}

\subsection{Dust Evolution Models}

Our models follow the radial evolution of a dust density distribution and calculate the growth, fragmentation, and erosion of dust particles, covering objects from 1\,$\mu$m to 2\,m in size. We do not take into account bouncing of particles. The sticking  probability depends on the relative velocities before collision. For this velocity, we take into account Brownian motion, vertical settling,  turbulent diffusion, and radial and azimuthal drift. For larger particles, the relative velocities increase \citep[e.g.][]{windmark2012}. The motion of the particles is determined by the interaction with the gas, and it is calculated according to their size \citep{birnstiel2010}.

We take parameters of a disk around a Herbig star, specifically  $T_{\star}=9300\,$K, $R_{\star}=2\,R_{\odot}$, and $M_{\star}=2.5\,M_{\odot}$. We assume that the gas surface density remains constant with time, which is assumed to be a power law with an exponential cutoff \citep[e.g.][]{lyndenbell1974}, that is,

\begin{equation}
\Sigma_g(r) = \Sigma_c \left(\frac{r}{r_c} \right)^{-\gamma} \times \exp\left[ -\left( \frac{r}{r_c}\right)^{2-\gamma} \right].
 \label{eq: sigma_gas}
\end{equation}

The cutoff radius $r_c$ is taken to be 120\,au, and $\gamma=1$. The total disk mass is $M_{\rm{disk}}=0.08\,M_{\odot}$. We assume a simple parameterization for the disk temperature that depends on the stellar parameters and is given by \citep{kenyon1987}

\begin{eqnarray}
T(r)&=&T_{\star}\left(\frac{R_{\star}}{r}\right)^{1/2} \phi_{\mathrm{inc}}^{1/4} \nonumber\\ \textrm{or} \qquad T(r)&\simeq&426\,\rm{K}\left(\frac{r}{1\,\rm{au}}\right)^{-1/2},
 \label{eq:temperature}
\end{eqnarray}

\noindent where $\phi_{\mathrm{inc}}$ is the angle between the incident radiation and the local disk surface, and it is taken to be $\phi_{\mathrm{inc}}=0.05$. We assume a logarithmically spaced radial grid from 1 to 500\,au with 500 steps. The initial gas-to-dust ratio is 100, and all the dust particles are assumed to be 1\,\micron~in size at the initial time.

We use simplified prescriptions of the fragmentation velocity threshold, which is assumed to change radially at the location of the ice lines corresponding to H$_2$O, NH$_3$, and CO$_2$. We assume that for grains whose mantle composition is dominated by nonpolar molecules (or with very low dipole moment), such as CO, CO$_2$, or silicates, the fragmentation velocity is of the order of 1\,m\,s$^{-1}$, in agreement with results from numerical simulations and laboratory experiments \citep{blum2000, poppe2000, paszun2009, gundlach2015, musiolik2016a, musiolik2016b}. Under the assumptions of our models, assuming the CO or CO$_2$ ice line does not make any difference, except that the location of the CO ice line is farther out, and $v_{\rm{frag}}$ would be 1\,m\,s$^{-1}$ for both ice lines. In this paper, we assume the CO$_2$ ice line, but results would be similar if we were to assume the CO ice line.

In the cases where the grain mantle is composed by molecules with permanent electrical dipoles, such as H$_2$O and NH$_3$, the fragmentation velocity is assumed to have a higher value. For  H$_2$O, it is assumed to be 10\,m\,s$^{-1}$ as suggested by numerical and laboratory experiments \citep{wada2009, wada2011, gundlach2015}. For NH$_3$, we assume a fragmentation velocity of $\sim$7\,m\,s$^{-1}$. This is an estimate, as there are no data on NH$_3$ fragmentation, but it is in correspondence with the contribution of the dispersion force to the total  Van der Waals force between dust grains, which is 24\% and 57\% for  H$_2$O and NH$_3$, respectively \citep{israelachvili1992}. For simplicity we assume the particles to be layered with a silicate core and mantles of water, ammonia, and carbon dioxide depending on their location relative to the ice lines.

At the ice lines of  H$_2$O, NH$_3$, and CO$_2$, we assume the fragmentation velocity to change as a smooth step function. We used a smooth step function, given by

\begin{equation}
 H(x) = \begin{dcases*}
        \frac{1}{2}\exp{\left(\frac{x}{\Delta x}\right)}  & for $x\leq 0$\\
        1- \frac{1}{2}\exp{\left(-\frac{x}{\Delta x}\right)}&for $x> 0$
        \end{dcases*}
\label{eq:Heaviside}
\end{equation}

\noindent where $x=r-r_{\rm{ice}}$, with $r_{\rm{ice}}$ being the radial position of a given ice line. These positions are assumed at the location where the disk temperature has the values of the average freezing temperatures of H$_2$O, NH$_{3}$, and CO$_2$, in agreement with the values reported in \cite{zhang2015} (Table~\ref{table_temperatures}). Notice that for CO$_2$, we take a freezing temperature of 44\,K, which is an averaged value for  CO$_2$ and CO. Our main motivation is to have this ice line at a distance of $\sim$90\,au, as observed in the disk around HD\,163296 (one of the targets that we use to compare our results with observations in Sect.~\ref{sect:comp_obs}). The factor $\Delta x$ in Eq.~\ref{eq:Heaviside} is a smoothing parameter for the radial change of the fragmentation velocity at a given position and is taken to be $\Delta x=0.5\,$au. Depending on how many ice lines are assumed, the fragmentation velocity is taken to be a composition of different smooth step functions (Eq.~\ref{eq:Heaviside}). For instance, for the simplest case, where only the water ice line is considered \citep[as in][]{birnstiel2010}, the fragmentation velocity is given by

\begin{equation}
v_{\rm{frag}}=\left(100*10^{H(x)}\right)\,\frac{\rm{cm}}{\rm{s}} \quad \mathrm{with} \quad x=r-r_{\rm{ice}, \rm{H}_2\rm{O}}
\end{equation}

%%%%%%%%%%%%
%TABLE 2
%%%%%%%%%%%%
\begin{table}
\label{table_temperatures}
\centering
\caption{Assumed freezing temperatures for H$_2$O, NH$_{3}$, and CO$_2$.}
\begin{tabular}{cccc}
\hline\hline
Parameter& H$_2$O & NH$_{3}$ & CO$_2$ \\
\hline
$T_{\rm{cond}}$ (K)&150&80&44 \\
$r_{\rm{ice}}$ (au)&8.1&28.5&92.0\\
\hline
\hline
\end{tabular}
\tablecomments{According to the averaged values reported in \cite{zhang2015} and based on \cite{mumma2011} and \cite{martin2014}.}
\end{table}

\begin{figure}
 \centering
   	\includegraphics[width=8.5cm]{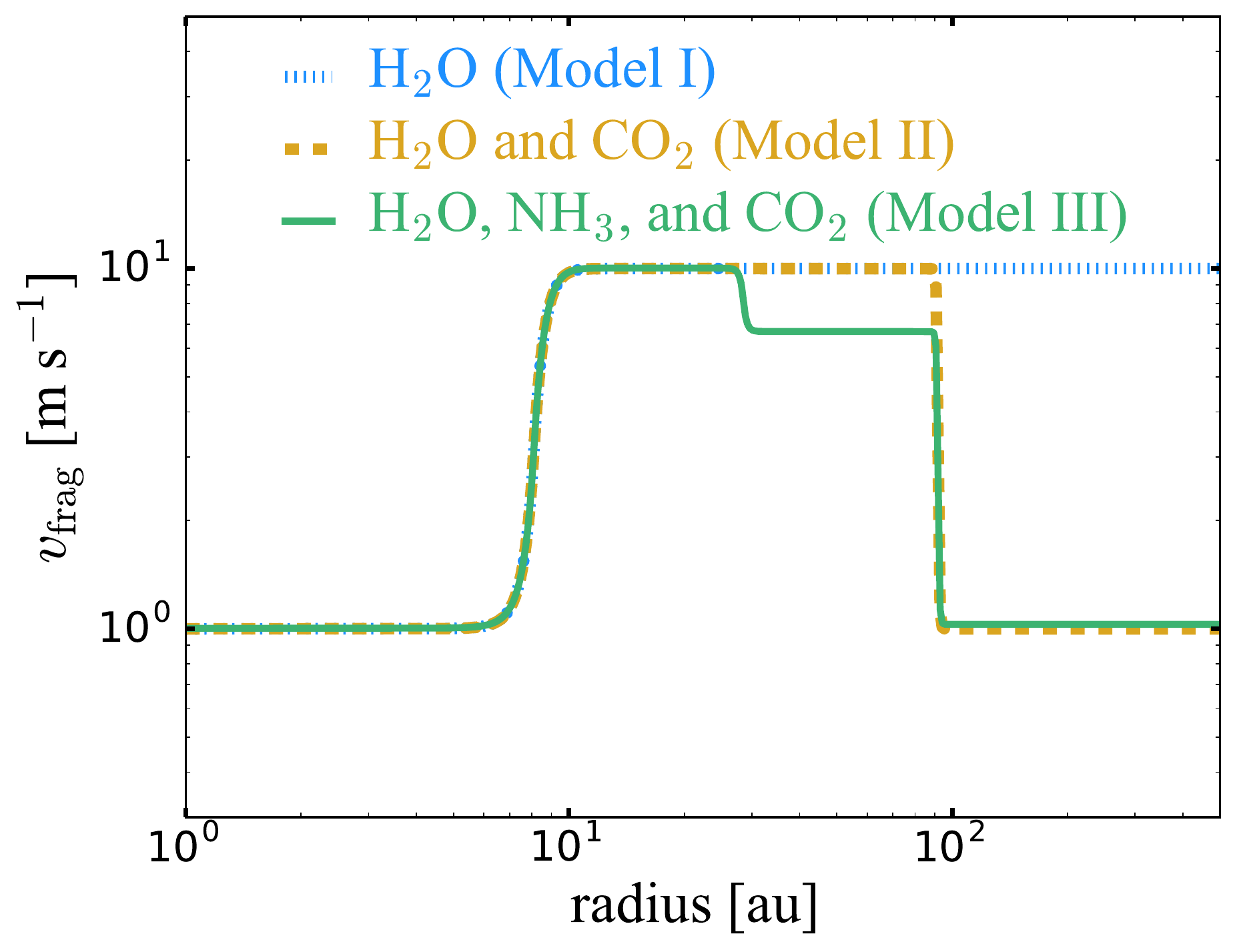}
   \caption{Profile of the fragmentation velocity for three different cases: (I) when only the H$_2$O ice line is assumed (dotted line); (II) when H$_2$O and CO$_2$ ice lines are assumed (dashed line); (III) when H$_2$O, NH$_3$, and CO$_2$ ice lines are assumed (solid line).}
   \label{fig:v_frag}
\end{figure}

In this work, we assume three different cases:  the fragmentation velocity only changes at H$_2$O ice line (model~I),  when the fragmentation velocity changes at the H$_2$O and CO$_2$ ice lines, being 1\,m\,s$^{-1}$ inside the  H$_2$O  ice line and beyond the  CO$_2$ ice line (model~II), and  when H$_2$O, NH$_3$, and CO$_2$ ice lines are assumed for the changes of the fragmentation velocity (model~III). Figure~\ref{fig:v_frag} shows the profiles of the fragmentation velocity for these three cases. With the disk and stellar properties of our models, the first case recreates the results already presented in \cite{banzatti2015}. In this paper, we add to this case the proper radiative transfer calculations to predict images at scattered-light and millimeter wavelengths.

In our models, the ice line locations and gas surface density remain constant with time. It is, however, expected that dust dynamics (in particular radial drift and vertical settling) can change the gas surface density and the location of ice lines \citep{piso2015, cleeves2016, krijt2016, powell2017, stammler2017}. In addition, the disk temperature can vary by different effects, such as disk dispersal, which can also change the ice line locations \citep{panic2017}. Our models are a simplification assuming that the viscous evolution timescales are longer than the dust growth timescales, and that any material that evaporates from the grains does not contribute significantly to the gas surface density, which is a reasonable assumption for the low dust-to-gas ratios that we have at any time of evolution in our models.

%This assumption implies that no pressure bumps are created at the location of any ice line, however, the amplitude of those pressure bumps are so low that dust trapping is not expected.

For the collisional outcome, the fragmentation probability  ($P_{\rm{frag}}$) is assumed to be unity when the relative velocity ($v_{\rm{rel}}$) is above the fragmentation velocity  ($v_{\rm{frag}}$) and zero when  $v_{\rm{rel}}$ is between 0 and 0.8\,$v_{\rm{frag}}$. For intermediate values (between 0.8 and $1\,v_{\rm{frag}}$), we assume a linear transition of the fragmentation probability, such that $P_{\rm{stick}}$=1-$P_{\rm{frag}}$ \citep{birnstiel2011}. When the collision velocity is only determined by turbulence, the maximum grain size is given by \citep{birnstiel2012}

\begin{equation}
	a_{\rm{frag}}=\frac{2}{3\pi}\frac{\Sigma_g}{\rho_s \alpha}\frac{v_{\rm{frag}}^2}{c_s^2},
  \label{eq:afrag}
\end{equation}

\noindent which is usually known as the fragmentation barrier. In Eq.~\ref{eq:afrag}, $\rho_s$ is the volume density of dust grains, which is taken to be $1.2\,\rm{g\,cm}^{-3}$ and $\alpha$ is a dimensionless quantity that it is typically used to parameterize the disk viscosity $\nu$ as \citep{shakura1973}

\begin{equation}
\nu=\alpha c_s  h \quad \textrm{with} \quad h=\frac{c_s}{\Omega},
\label{eq:viscosity}
\end{equation}

\noindent where $\Omega=\sqrt{GM_\star r^{-3}}$ ($G$ is the gravitation constant) and $c_s$ is the sound speed.  Particles can reach large sizes drifting toward the star in timescales shorter than the collision timescales (mainly in the other parts of the disk). This barrier to further growth is the drift barrier, and it is given by

\begin{equation}
	a_{\rm{drift}}=\frac{2 \Sigma_d}{\pi\rho_s}\frac{v_K^2}{c_s^2}\left \vert \frac{d \ln P}{d\ln r} \right \vert^{-1},
  \label{eq:adrift}
\end{equation}

\noindent where $\Sigma_d$ is the dust density distribution, $v_K$ is the Keplerian angular velocity (i.e. $v_K=r\Omega$), and $P$ is the disk pressure $P=\rho c_s^2$, with $\rho$ being the total gas density.

\cite{okuzumi2016} modeled the dependency of the fragmentation velocity of dust particles on the sintering. We do not take this particle fusion process into account, which can also affect the fragmentation velocity and create gaps and rings in the dust surface density.

%%%%%%%%%%%%
%FIGURE
%%%%%%%%%%%%

\subsection{Radiative Transfer Models} \label{RT_models}

For the radiative transfer calculations, we follow the same procedure presented in \cite{pohl2016}. However, for the near-infrared predictions, we do not calculate images in the full Stokes vector as in \cite{pohl2016}, but concentrate on the total intensity. In addition, we assume the temperature as in the dust evolution models. The models are 3D, but we treat the azimuth with only one grid cell since our dust evolution models are azimuthally symmetric. The total dust density is calculated assuming the dust density distribution from the dust evolution models, such that

\begin{equation}
	\rho_{\mathrm{d}}(R,\varphi,z) = \frac{\Sigma_{\mathrm{d}}(R)}{\sqrt{2\,\pi}\,H_{\mathrm{d}}(R)}\,\exp \left( -\frac{z^2}{2\,H_{\mathrm{d}}^2(R)} \right)\,,
	\label{eq:volume_density}
\end{equation}

\noindent where $R$ and $z$ are spherical coordinates, that is, $R = r\,\sin(\theta)$ and $z = r\,\cos(\theta)$, with $\theta$ being the polar angle. In order to include the effect of settling on the models, the  dust scale height $H_{\mathrm{d}}(R)$ depends on the grain size and the disk viscosity \citep{youdin2007, birnstiel2010}, and it is given by

\begin{equation}
	H_{\mathrm{d}}=h \times \mathrm{min} \left( 1,\sqrt{\frac{\alpha}{\mathrm{min}(\mathrm{St},1/2)(1+\mathrm{St}^2)}}\,\right)
	\label{eq:dust_scaleheight}
\end{equation}

\noindent where $h$ is the gas scale height (Eq.~\ref{eq:viscosity}) and St is the Stokes number, which is a parameter that quantifies the coupling of the particles on to the gas; and at the midplane is given by

\begin{equation}
\textrm{St}=\frac{a\rho_s}{\Sigma_g}\frac{\pi}{2}.
\label{eq:stokes}
\end{equation}

%%%%%%%%%%%%
%FIGURE
%%%%%%%%%%%%
\begin{figure*}
 \centering
   	\includegraphics[width=18cm]{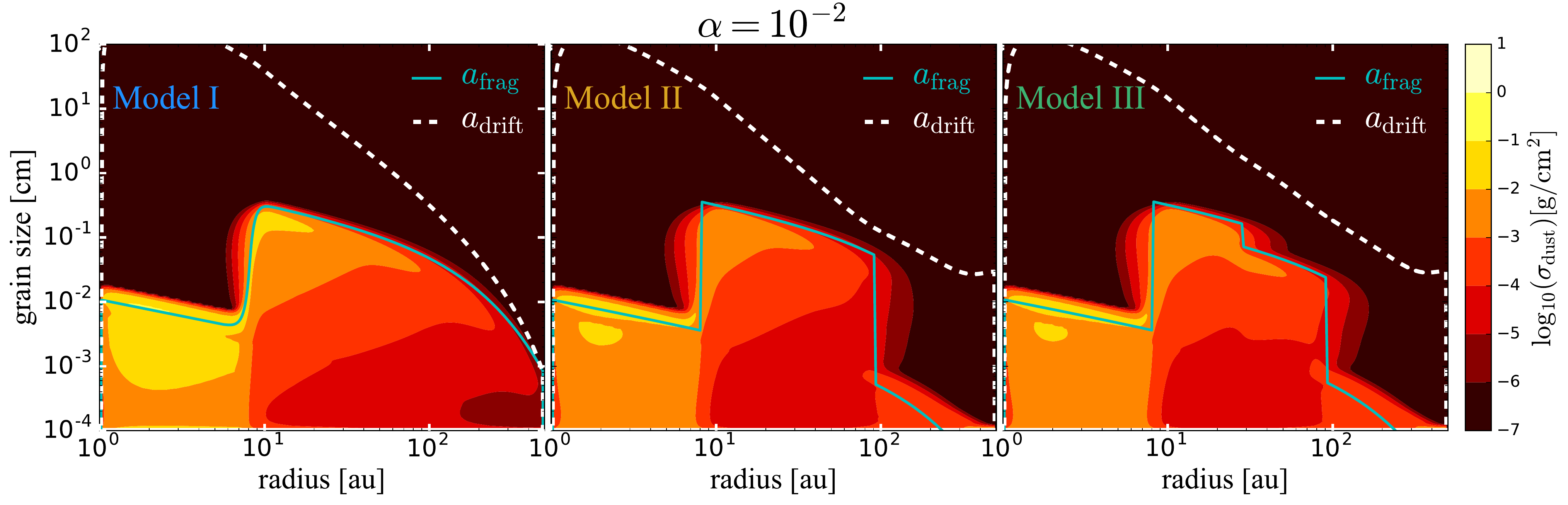}\\
	\includegraphics[width=18cm]{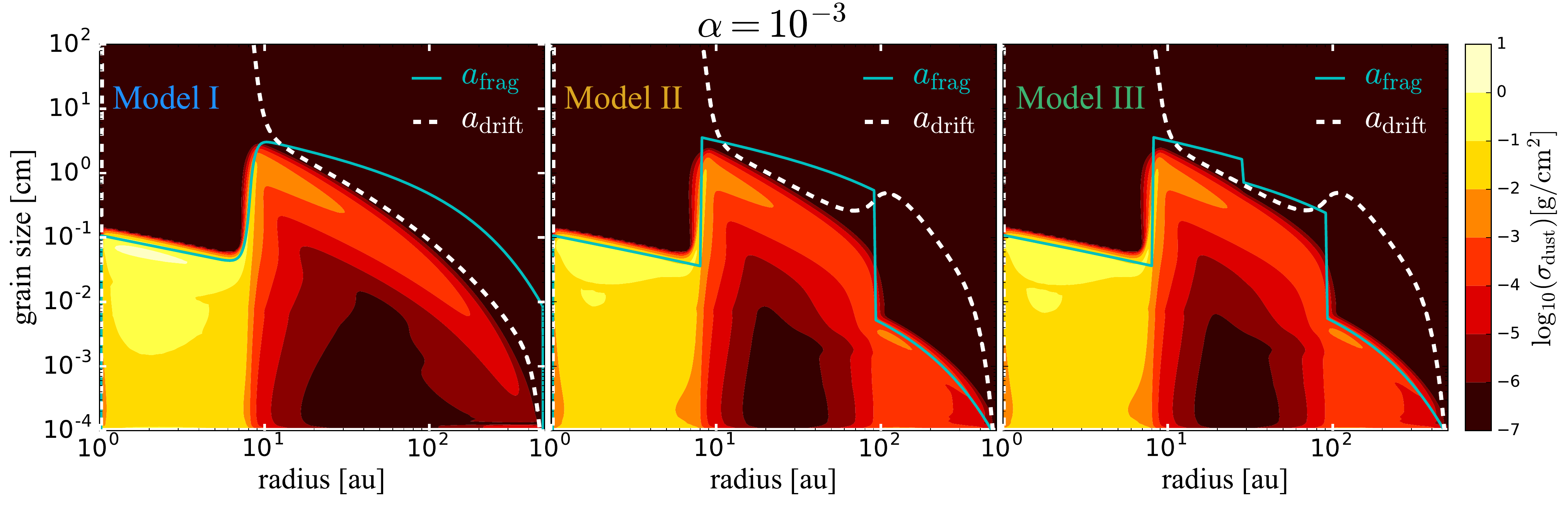}\\
	\includegraphics[width=18cm]{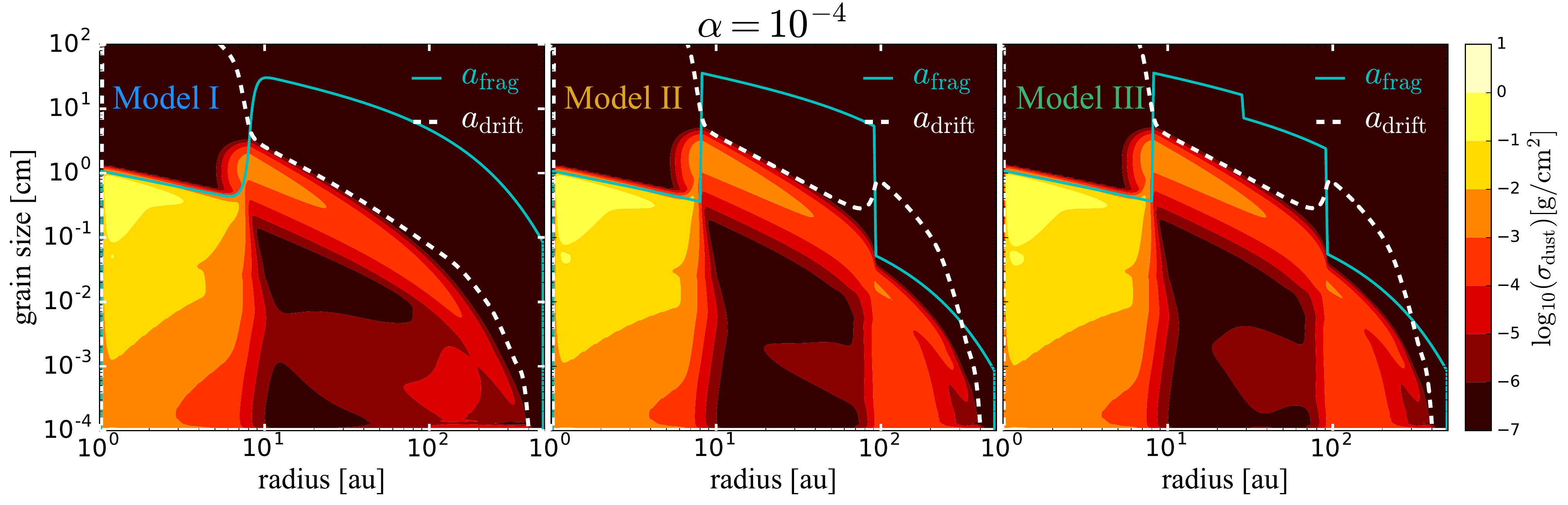}
   \caption{Vertically integrated dust density distribution at 1\,Myr of evolution when one, two, or three ice lines are considered (see Fig.~\ref{fig:v_frag}, for reference). Results are for $\alpha=10^{-2}$ (top panels),  $\alpha=10^{-3}$ (middle panels), and  $\alpha=10^{-4}$ (bottom panels). The solid cyan line represents the maximum grain size set by fragmentation (Eq.~\ref{eq:afrag}), and the dashed white line corresponds to the maximum grain size set by radial drift (Eq.~\ref{eq:adrift}). }
   \label{fig:dust_density}
\end{figure*}

For the synthetic images, we assume a distance to the disk of 140\,pc and a face-on disk. For the convolved synthetic images, we assume a Gaussian for the point spread function (PSF) with an FWHM of 0.\arcsec04 for the total intensity at 1.6\,$\mu$m, as a realistic value for observations with VLT/SPHERE or GPI. For images at the continuum millimeter emission, we consider two wavelengths, 0.87 and 3.0\,mm, and we convolve the images with a Gaussian beam of 0.\arcsec04, in order to have the same resolution as at short wavelength and to mimic ALMA observations with high resolution.

For the dust optical properties, we assume a dust composition as in \cite{ricci2010}, which is a mixture between silicate, carbonaceous, and water ice. In \cite{banzatti2015}, we investigated how different fractions of water ice at the snow line could affect the resulting dust density distributions and hence the observational predictions. We found a weak effect on the results, and for this reason in this paper we keep the dust composition fixed for all the models.

\section{Results}

\subsection{Dust Density Distributions} \label{sect:results}

Figure~\ref{fig:dust_density} shows the vertically integrated dust density distribution at 1\,Myr of evolution when one, two, or three ice lines are considered (see Fig.~\ref{fig:v_frag}, for reference). Results are shown for three different values of  $\alpha$ viscosity, specifically $\alpha=[10^{-4},10^{-3},10^{-2}]$. Figure~\ref{fig:dust_density2} shows the dust density distribution for small ($a\in[1-10]\,\mu$m) vs. large grains ($a\in[1-10]\,$mm)  for the same cases as Fig.~\ref{fig:dust_density}. In this section, we describe the radial variations of the dust density distribution for each value of $\alpha$ viscosity. In the Appendix, the vertical dust density distribution (Eq.~\ref{eq:volume_density}) is shown for each case. These distributions are assumed for the radiative transfer calculations.

\subsubsection{Case of $\alpha=10^{-2}$}

In the case that the viscosity is high, the maximum grain size is determined by fragmentation, and in this case $a_{\rm{frag}}<a_{\rm{drift}}$ for the entire disk. As a consequence, the radial changes of $v_{\rm{frag}}$ have a direct effect on the resulting dust density distributions (Eq.~\ref{eq:afrag}).

In the case when only the water ice line is included (model I, Fig.~\ref{fig:v_frag}),  the maximum grain size decreases smoothly with radius beyond the water ice line (Eq.~\ref{eq:afrag}), where the maximum grain size is around 3\,mm. In the outer disk, the distribution of small grains is as extended as the gas density. Due to the effective destructive collisions when the fragmentation velocity is taken to be 1\,m\,s$^{-1}$, that is, for $r\lesssim r_{\rm{ice}, \rm{H}_2\rm{O}}$, there is an enhancement of the dust density for small grains (Fig.~\ref{fig:dust_density} and Fig.~\ref{fig:dust_density2}) within this region. In these inner locations, the maximum grain size is around 0.1\,mm.  Hence, the surface density of millimeter-sized particles shows a depletion for $r\lesssim r_{\rm{ice}, \rm{H}_2\rm{O}}$, because particles do not grow to these large sizes. However, outside the water ice line, where the fragmentation velocity is higher (10\,m\,s$^{-1}$), the dust density distribution for millimeter grains increase from $\sim8\,$au up to 30\,au. At $\sim$30\,au the dust density of millimeter-sized particles starts to decrease outward. The outer radius for the dust density of the millimeter grains is around 70\,au (we call the outer radius the location where the dust density drops lower than $10^{-7}$g\,cm$^{-2}$, Fig.~\ref{fig:dust_density2}), while the gas outer radius is 500\,au (the cutoff radius $r_c$ is at 120\,au). This is because of radial drift as shown by \cite{birnstiel2014}.

When the H$_2$O and CO$_2$ ice lines are both included for the calculation of $v_{\rm{frag}}$ (model II, Fig~\ref{fig:v_frag}), there is a re-creation of small grains at  $r\gtrsim r_{\rm{ice}, \rm{CO}_2}$, and therefore there is an enhancement of the density of the small grains from this location. As a consequence, a gap-like shape in the dust density distribution of small grains is formed within the region between the ice lines  of H$_2$O and CO$_2$ (Fig.~\ref{fig:dust_density}) and surrounded by two ring-like shapes in the dust surface density distribution of the micron-sized particles. In this case, the dust density decreases by $\sim1-2$ orders of magnitude for small grains (Fig.~\ref{fig:dust_density2}), while the dust density for millimeter-sized particles is enhanced within the same region. The outer radius of the dust density of the millimeter grains is slightly farther out in this case than in model~I because at the same time of evolution (1\,Myr), there is a higher replenishment of small grains in the outer part of the disk, for which the drift velocities are lower, and it takes longer times for the particles to reach the locations where they can grow to larger sizes, that is, $r \lesssim r_{\rm{ice}, \rm{CO}_2}$.

The inclusion of the NH$_3$ ice line does not have a significant effect on the final dust density distribution compared to model II (Fig~\ref{fig:v_frag}). In this case the fragmentation velocity decreases at $r_{\rm{ice}, \rm{NH}_3}$ from 10 to 7\,m\,s$^{-1}$, and there is a more effective fragmentation  at $r_{\rm{ice}, \rm{NH}_3}$  than in model II from $r_{\rm{ice}, \rm{NH}_3}$. This is reflected in a small increment of the dust density of small grains at such locations  (Fig.~\ref{fig:dust_density2}). The shape of the dust density distribution for large millimeter-sized particles remains similar to that in models I and II.

In summary, for this case of $\alpha=10^{-2}$, $a_{\rm{max}}=a_{\rm{frag}}$ for the entire disk and variations of $v_{\rm{frag}}$ affect the final distribution of small and large particles. In the region between the H$_2$O and CO$_2$ ice lines, the density of small grains is depleted by $\sim2$ orders of magnitude compared to the inner disk. In the outer disk, $r\gtrsim r_{\rm{ice}, \rm{CO}_2}$, the density of small grains increases owing to the effective fragmentation of particles in this region. The ice line of NH$_3$ has little effect on the distribution of small grains. In all models, the growth due to the sticking properties of particles leads to a distribution of the millimeter-sized particles with a ring-shaped structure in the region between the H$_2$O and CO$_2$.

\subsubsection{Case of $\alpha=10^{-3}$}

In the case where the viscosity has an intermediate value, there are regions in the disk where the maximum grain size is determined by fragmentation and where it is determined by drift (Fig.~\ref{fig:dust_density}). Because $\alpha$-viscosity is lower, particles can reach larger sizes (Eq.~\ref{eq:afrag}).

%%%%%%%%%%%%
%FIGURE
%%%%%%%%%%%%
\begin{figure*}
 \centering
   	\includegraphics[width=18cm]{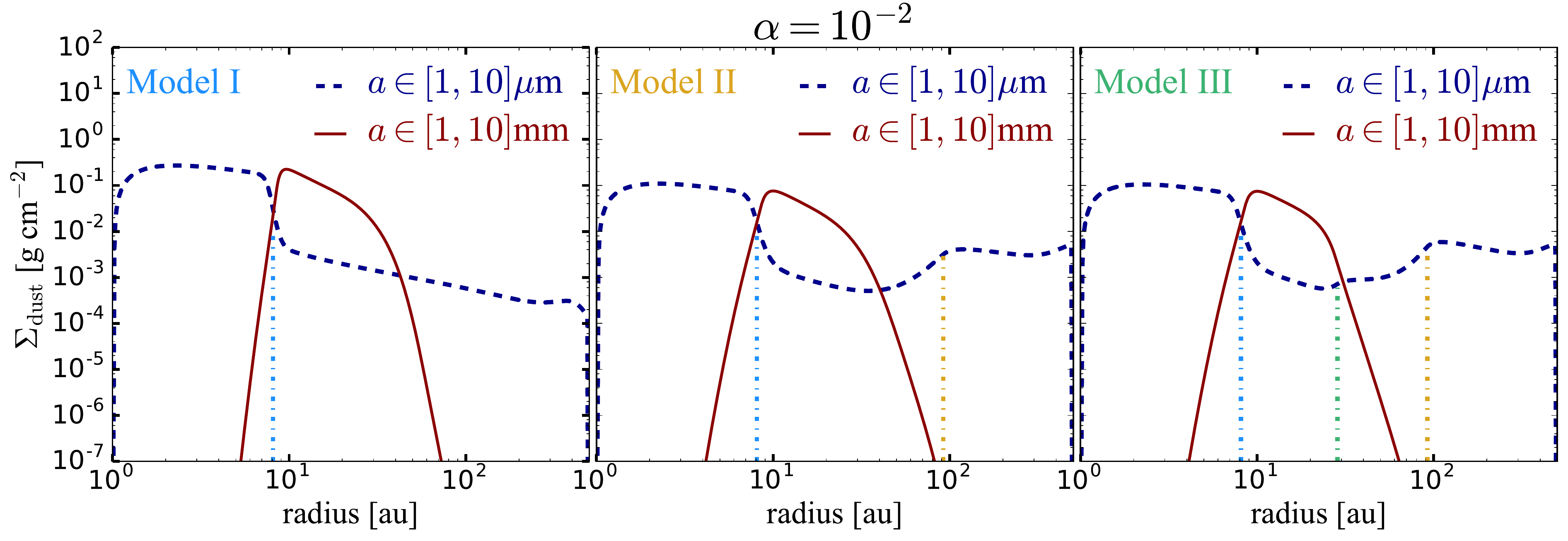}\\
	\includegraphics[width=18cm]{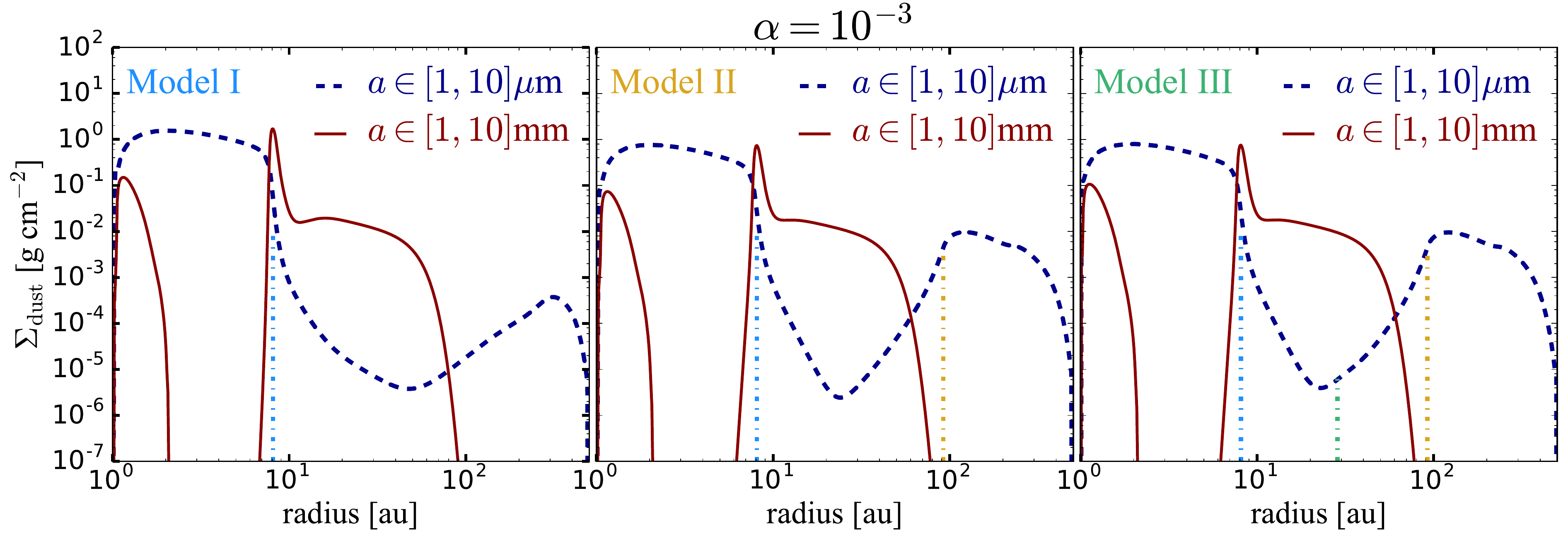}\\
	\includegraphics[width=18cm]{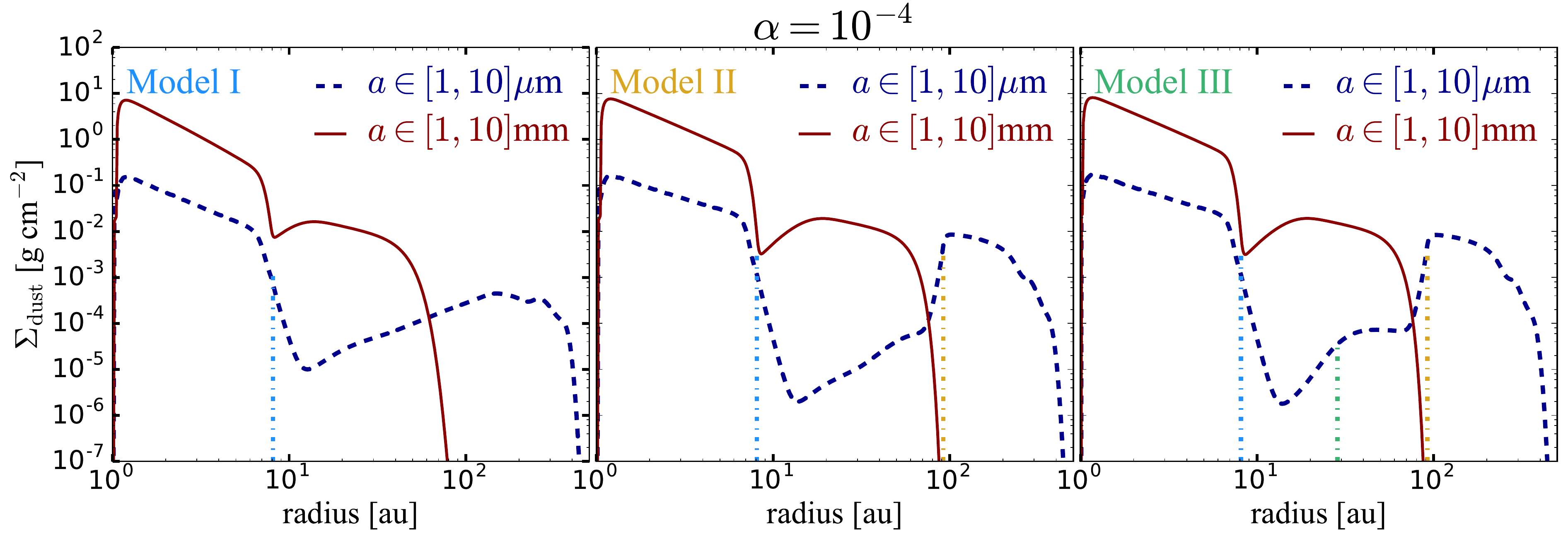}
   \caption{Dust density distribution for small ($a\in[1-10]\,\mu$m) vs. large grains ($a\in[1-10]\,$mm)  at 1\,Myr of evolution when one, two, or three ice lines are considered (see Fig.~\ref{fig:v_frag} for reference, and vertical lines that represent the ice line locations according to Table~\ref{table_temperatures}). Results are for $\alpha=10^{-2}$ (top panels),  $\alpha=10^{-3}$ (middle panels), and  $\alpha=10^{-4}$ (bottom panels). }
   \label{fig:dust_density2}
\end{figure*}

In model I, inside the water ice line $a_{\rm{max}}=a_{\rm{frag}}$ as in the case of $\alpha=10^{-2}$. The maximum grain size in this region is $\sim1\,$mm (one order of magnitude higher than in the case of $\alpha=10^{-2}$), but it remains depleted of larger grains, and there is an enhancement of the distribution of small grains as before. For $r\gtrsim r_{\rm{ice}, \rm{H}_2\rm{O}}$,  $a_{\rm{max}}$ is determinated by drift, the growth is more effective, and particles reach  sizes of 1\,mm$\lesssim a \lesssim $3\,cm,  depleting this region in small grains. Because grains grow to larger sizes in $r\gtrsim r_{\rm{ice}, \rm{H}_2\rm{O}}$, they are in the drift limit, that is, the maximum grain size is determined by Eq.~\ref{eq:adrift}. At $r_{\rm{ice}, \rm{H}_2\rm{O}}$, there is a narrow ring-like accumulation of millimeter grains, where $v_{\rm{frag}}$ changes from 1 to 10\,m\,s$^{-1}$. This happens just at the location where $a_{\rm{frag}}<a_{\rm{drift}}$ (Fig~\ref{fig:dust_density} and \ref{fig:dust_density2}). Beyond this peak of the density of large grains, the distribution of the millimeter-sized particles is uniform up to 60\,au where the dust density sharply decreases with radius as in the case of $\alpha=10^{-2}$. For this case, the dust density distribution of large grains reassembles similar profiles to those in the case of dust trapping by a broad pressure bump \citep[e.g. created by a planet or at the outer edge of a dead zone;][]{pinilla2012, pinilla2016}. In model~I, the small grains also show a gap in the outer regions, which is a result of inefficient fragmentation of dust particles as shown by \cite{birnstiel2015}.

In model II, due to the changes of $v_{\rm{frag}}$ at $r_{\rm{ice}, \rm{CO}_2}$, the maximum grain size in the outer parts of the disk ($r\gtrsim r_{\rm{ice}, \rm{CO}_2}$) is again determined by fragmentation. In this region there is effective fragmentation of particles, and the dust density distribution of small grains again increases in the outer disk (Fig~\ref{fig:dust_density2}), creating a clear gap  where the dust density for $a\in[1-10]\,\mu$m decreases around 4 orders of magnitude. The distribution of large grains does not significantly change compared to model~I. There is an accumulation of large grains at the location of the water ice line and then a uniform distribution up to 60\,au, where the density decreases sharply with radius. The small bump at the location of the water ice line is a traffic jam effect where particles, for which maximum size is determined by radial drift, reduce their radial velocities because they fragment, or, in other words, when their size decreases and their drift velocities are lower.

In model III, though there is a change of $v_{\rm{frag}}$ at $r_{\rm{ice}, \rm{N}_3\rm{H}}$, there is no effect on the final dust density distribution (Fig~\ref{fig:dust_density}). This is because at these locations $a_{\rm{max}}=a_{\rm{drift}}$ and hence changes of $v_{\rm{frag}}$ do not influence the distribution of small grains as in the case of model III with $\alpha=10^{-2}$. The distribution of  large grains is similar to that in models I and II.

Summarizing, for this case of $\alpha=10^{-3}$,  in all models (I, II, and III) there is a narrow ring-like accumulation of millimeter-sized particles in the dust density distribution at  $r_{\rm{ice}, \rm{H}_2\rm{O}}$ where $v_{\rm{frag}}$ changes from 1 to 10\,m\,s$^{-1}$; beyond that accumulation the distribution of large grains is uniform and the outer radius for the large grains is around 80-90\,au. For small grains, there is a decrease of the distribution of small grains due to inefficient fragmentation. When $v_{\rm{frag}}$ changes again to 1\,m\,s$^{-1}$ in the outer disk, because CO$_2$ ice behaves like silicates in terms of collisions, there is an enhancement for the dust distribution of small particles, creating a distinct gap. The inclusion of the NH$_3$ ice line, do not change the final distribution of particles.

\subsubsection{Case of $\alpha=10^{-4}$}

In the case where the viscosity has a low value, there are regions in the disk where the maximum grain size is determined by fragmentation and where it is determined by drift as in the case of $\alpha=10^{-3}$. Because the viscosity is lower, $a_{\rm{frag}}$ increases by one order of magnitude, and thus the maximum grain size, in the regions where $a_{\rm{max}}=a_{\rm{frag}}$. This is the case inside the water ice line, where  $a_{\rm{max}}$ is around 1\,cm. This implies that there is no depletion of millimeter-sized grains inside the water ice line because $a_{\rm{frag}}$ is large enough to allow millimeter grains, due to the low $\alpha$ (Fig.~\ref{fig:dust_density}). On the contrary, there is an enhancement of large ($a\in[1,10]\,$mm) particles inside the water ice line. An enhancement of the dust surface density beyond the water ice line only occurs for the very large grains ($a>1$\,cm; Fig.~\ref{fig:dust_density}).

In model I, beyond the water ice line $a_{\rm{max}}=a_{\rm{drift}}$ and the maximum grain size only increases by a small factor. Although the particles reach larger sizes, there is not a broad ring-like accumulation of grains with size of 1-10\,mm as in the case of $\alpha=10^{-3}$. Instead, there is a small depletion of the millimeter-sized grains at $r_{\rm{ice}, \rm{H}_2\rm{O}}$ as for the smallest grains (Fig.~\ref{fig:dust_density}). This is because fragmentation is too inefficient in this region, creating gaps for the dust surface density of large and small grains. Beyond this gap, there is a uniform distribution of millimeter-sized particles. In this case, the dust diffusion is slower because the viscosity is lower, and therefore it takes longer times for the small particles to diffuse outward. As a result, the gap is slightly narrower in small grains than in the previous case of $\alpha=10^{-3}$.

In model II, as in the previous cases, due to the changes of $v_{\rm{frag}}$, $a_{\rm{max}}=a_{\rm{frag}}$ for locations beyond the CO$_2$ ice line. This makes the gap of the small grains deeper, with a depletion factor of around 4 orders of magnitude.  As in the previous case, the distribution of millimeter-sized particles does not remarkably change between models I and II.

In model III, the distribution of small and large particles is similar to that in models I and II. There is a small enhancement in the distribution of small particles close to $r_{\rm{ice}, \rm{NH}_3}$. This is the result of the slower diffusion of small grains, which take longer times to diffuse outward. In this case, at around 1.5\,Myr, this enhancement disappears.

In summary, the dust density distributions when $\alpha=10^{-4}$ are similar to those when $\alpha=10^{-3}$. The main difference lies at the location of the water ice line, where there is a gap of millimeter grains contrary to the case of $\alpha=10^{-3}$, where there is a small bump. This difference is due to the inefficient  fragmentation at this region, which depletes the region of micron- and millimeter-sized particles simultaneously.

\begin{figure*}
 \centering
   	\includegraphics[width=18cm]{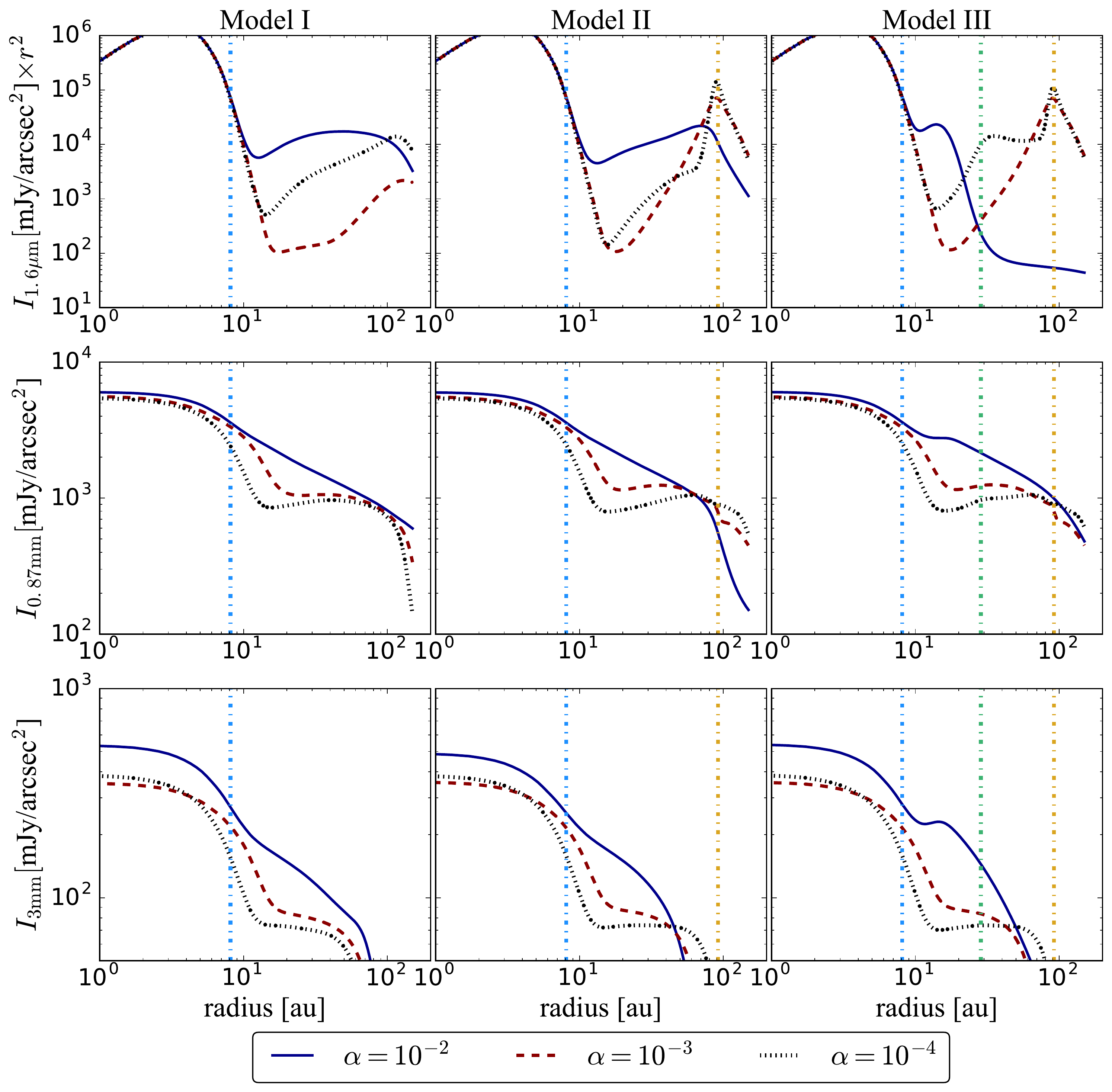}
    \caption{Radial intensity profile of the synthetic images after radiative transfer calculations assuming the dust density distributions from Fig.~\ref{fig:dust_density}. The intensity profiles are obtained at 1.6\,$\mu$m (top panels, which are multiplied by $r^2$), 0.87\,mm (middle panels), and 3\,mm (bottom panels), for the cases where one, two, or three ice lines are considered (from left to right panels), and assuming different values for $\alpha$. The vertical lines represent the ice lines (Table~\ref{table_temperatures}).}
   \label{radial_profile_synthetic}
\end{figure*}

\subsection{Radial Intensity Profiles} \label{sect:images}
\begin{figure*}
 \centering
   	\includegraphics[width=18cm]{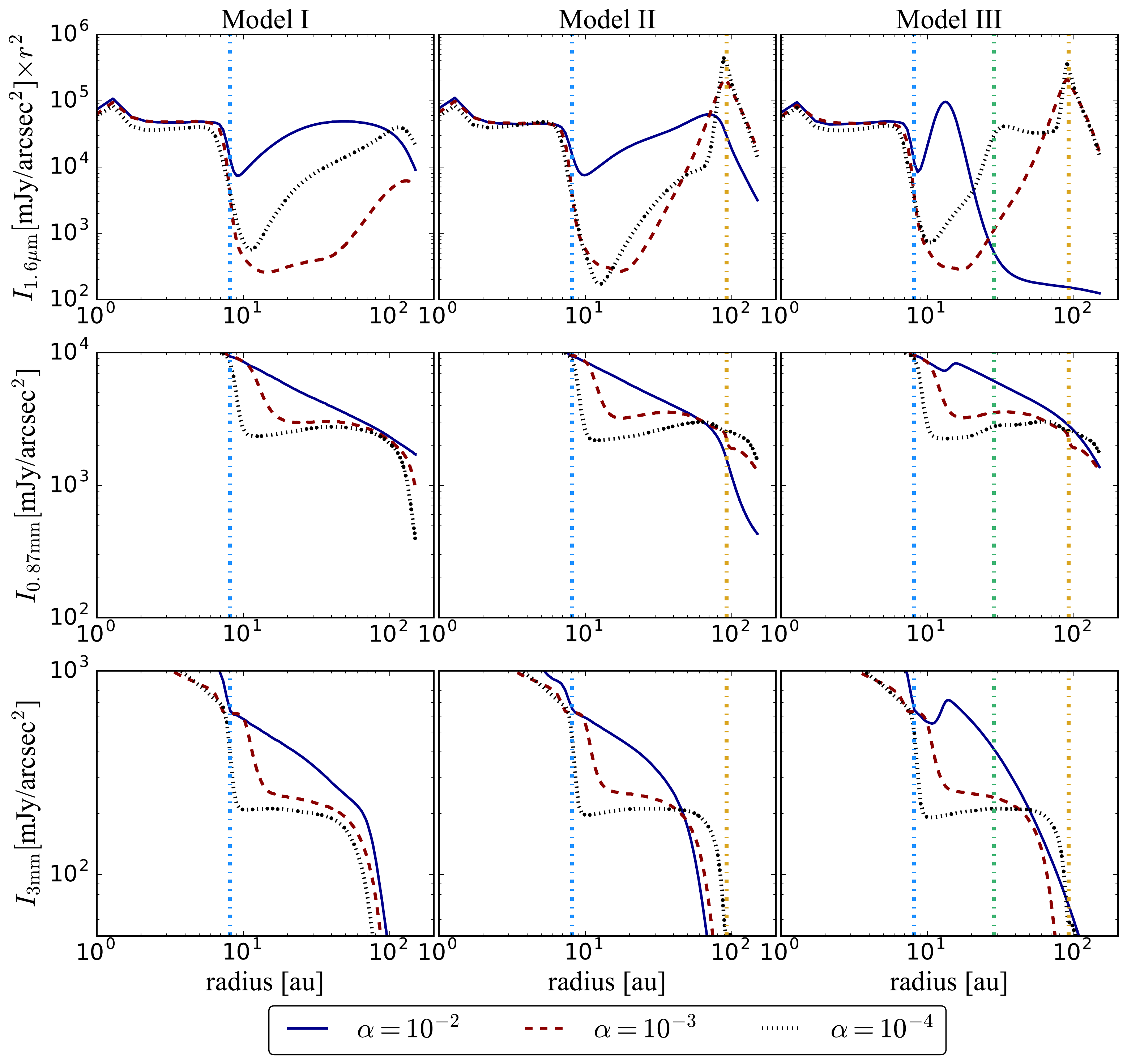}
    \caption{Same as Fig.~\ref{radial_profile_synthetic,} but convolved with a circular PSF with an FHWM of 0.\arcsec04 for all profiles.}
     \label{radial_profile_convolved}
\end{figure*}

After the radiative transfer calculations described in Sect~\ref{RT_models}, we obtain images at near infrared and (sub) millimeter emission, specifically at 1.6\,$\mu$m, 0.87\,mm, and 3\,mm. Figure~\ref{radial_profile_synthetic} shows the radial profiles of the intensity from the synthetic images. Figure~\ref{radial_profile_convolved} shows the same profiles after convolving with a circular PSF with a FHWM of 0.\arcsec04 (which corresponds to 5.6\,au for the distance that we assume of 140\,pc), in order to compare the multi-wavelength profiles at the same spatial resolution. The intensity profiles at 1.6\,$\mu$m are multiplied by $r^2$, to compensate for the $r^{-2}$ dependency of the stellar illumination.

The intensity profiles reveal different kinds of structures depending on the assumed value for $\alpha$ viscosity and on the number of the ice lines that are assumed in the models. In our models the radial structures at scattered-light and millimeter emission originate from  variations (radial and vertical; see figures in Appendix) of the total dust density distribution, and not from temperature variations. The temperature that we assume is a simple power law (Eq.~\ref{eq:temperature}), as in the dust evolution models. As a test, we checked whether the obtained intensity profiles change when the temperature is also calculated in the radiative transfer models, and we do not find significant changes. This is because the temperature values are not very different from the power-law assumption, and the only significant variations are in the very inner disk ($\sim1$\,au), where the maximum temperature is higher from the radiative transfer calculations.

When only the water ice line is assumed (model~I), the general trend is a dip of emission near the location of the water ice line at all different wavelengths. The dip is deeper at 1.6\,$\mu$m than at (sub)millimeter emission, and for lower viscosity the depression of the dip is higher. These results are similar to those of \cite{birnstiel2015}, who found dips of emission in the outer parts of protoplanetary disks at scattered light due to inefficient fragmentation of dust particles. The main difference with the present models is that the dip is located closer in, near the water ice line, where the fragmentation velocity is expected to change, and not in the outer disk. Although the resulting dust surface density distribution in model~I shows a depletion of millimeter grains inside the water ice line and a ring-like structure just beyond, the intensity at (sub)millimeter emission does not show such ring-like structure, and instead it also shows a dip of emission. This is because inside the water ice line there is an enhancement of all grain sizes below the maximum grain size, which is determined by fragmentation $a_{\rm{frag}}$. All these grains contribute to the emission at (sub)millimeter wavelengths, and when fragmentation becomes less efficient beyond the water ice line, there is a depression in the total intensity because grain growth and drift are more efficient. This is opposite to the case of particle trapping by a planet or at the outer edge of a dead zone, where only the dust grains inside the trap emit at (sub)millimeter wavelengths.

For models~II and III, when CO$_2$ and NH$_3$ ice lines are included in addition to the water ice line, there are  changes in the radial profile of the intensity. The main difference is for the intensity at 1.6\,$\mu$m, where a clear gap of emission is formed, becoming deeper when $\alpha$ has values of $10^{-3}$ and $10^{-4}$. At the outer edge of that gap, a ring-like structure is also formed, and in the case of three ice lines and low viscosity, there are multiple ring- and gap-like structures. For the case of $\alpha=10^{-2}$, the gap in micron-sized particles is shallower than in the other two cases because when $\alpha$  decreases, the growth is more efficient between the water and the other ice lines. As a result, there is a high depletion of micron-sized particles, creating a deep gap in the intensity profile, whose depletion factor is around four orders of magnitude when comparing to the inner part of the disk.  The (sub)millimeter emission in the case of multiple ice lines remains similar to that in the case when only the water ice line is included. This is because the millimeter/centimeter-sized particles are not affected by the changes of the fragmentation velocity when more than one ice line is included in the outer disk. In model~III and when $\alpha=10^{-2}$, there is a sharp decrease of the intensity at 1.6\,$\mu$m after the gap and ring located near the water ice line. This is the result of a more flared inner disk with high density of small grains that can block the light from the central star, preventing the light from being scattered in the outer surface layers.

In summary, the general trend in all models is just a dip of emission (or shallow gaps) at (sub)millimeter emission near the ice lines. The width of the gap or the separation between the ring-like shapes is smaller at longer wavelengths. The depression of such dip or gap at (sub)millimeter is less than one order of magnitude and thus much more shallower than at 1.6\,$\mu$m. It is important to note that at longer wavelengths, that is 3\,mm, the outer disk radius is smaller than at shorter wavelengths as expected from radial drift.

In the convolved images the structures do not show a significant change compared to the theoretical images, and they prove that the structures and depletions that are expected from ice lines can be detected with the current capabilities of different telescopes, such as ALMA, VLT/SPHERE, and GPI.

\vspace{0.5cm}

\section{Discussion} \label{sect:discussion}
\subsection{The Effect of the Disk Viscosity}

The assumed value  for the disk viscosity ($\alpha$) plays an important role in the final dust distributions and hence in the  potential structures. The value of $\alpha$ depends on how angular momentum is transported within the disk, which can have different origins such as MRI and magnetohydrodynamical (MHD) winds \citep[e.g.][]{balbus1991, suzuki2009, bai2016}. MRI requires disk ionization for the disk gas to be coupled to the magnetic field, and therefore if MRI drives accretion and turbulence, the value of $\alpha$ depends on the ionization environment of the disk \citep[e.g.][]{dolginov1994, flock2012, desch2015}.

Observationally, measuring the level of turbulence in protoplanetary disks is quite challenging and remains very uncertain, although recent efforts with ALMA observations have allowed us to measure the turbulent velocity dispersion for a couple of protoplanetary disks \citep{flaherty2015, flaherty2017, teague2016}, revealing values predicted by MRI, but still with large uncertainty \citep[$\alpha\sim10^{-4}-10^{-2}$; e.g.,][]{simon2015}.

The value of $\alpha$ is of great importance in the context of dust evolution models because it determines the turbulent velocities of the dust particles and hence the fragmentation limit (Eq.~\ref{eq:afrag}). After combining our results from the dust evolution models with radiative transfer calculations, we demonstrate that the shape (in particular the depth) of the dips or gaps that formed due to variations of fragmentation dust properties near the ice line depends on $\alpha$. The depletion factor of the gaps or dips becomes higher for lower values of $\alpha$, although the main differences are between the results of $10^{-2}$ and the other two values considered ($10^{-4}$ and $10^{-3}$). This is because for $\alpha=10^{-2}$, the maximum grain size is determined by fragmentation in the entire disk, while for the other two values it is a mix between fragmentation and drift. Therefore, observational insights about the turbulence in disks can provide better constraints on whether or not the origin of gaps is due to ice lines.

In the context of our simulations, the number of rings or gaps depends on how many changes of the fragmentation velocity are assumed. In this paper, we assume up to three variations corresponding to three main volatiles. It is important to notice that multiple gaps/rings are only obtained at the near-infrared emission, while at the millimeter emission there is only a dip or a single gap. Multiple ring structures at scattered light were only obtained in model~III with three radial variations of the fragmentation velocity and low values of $\alpha-$viscosity (Fig.~\ref{radial_profile_synthetic} and Fig.~\ref{radial_profile_convolved}).

\begin{figure*}
 \centering
   \begin{tabular}{cc}   
   	\includegraphics[width=8.5cm]{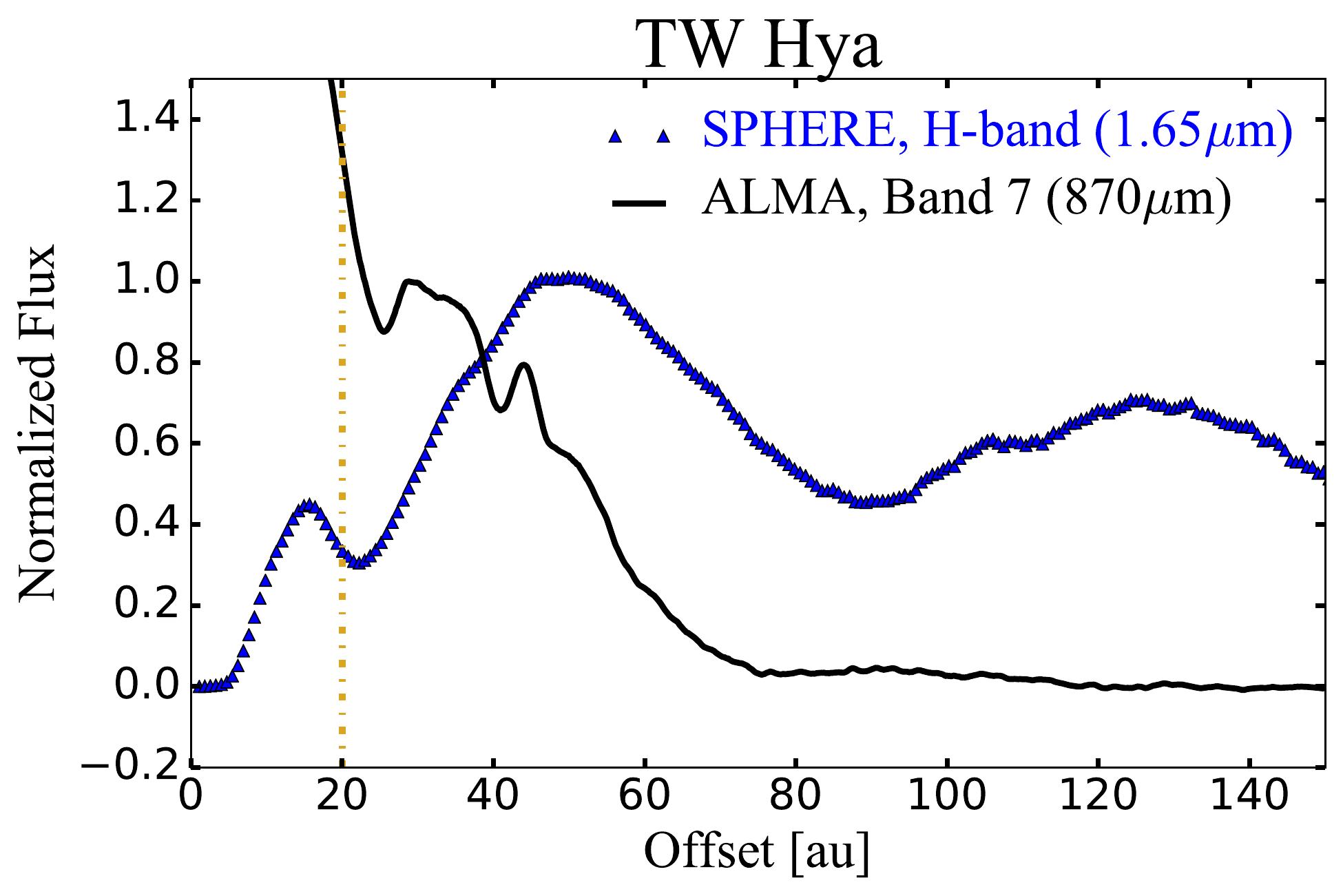}&
	\includegraphics[width=8.5cm]{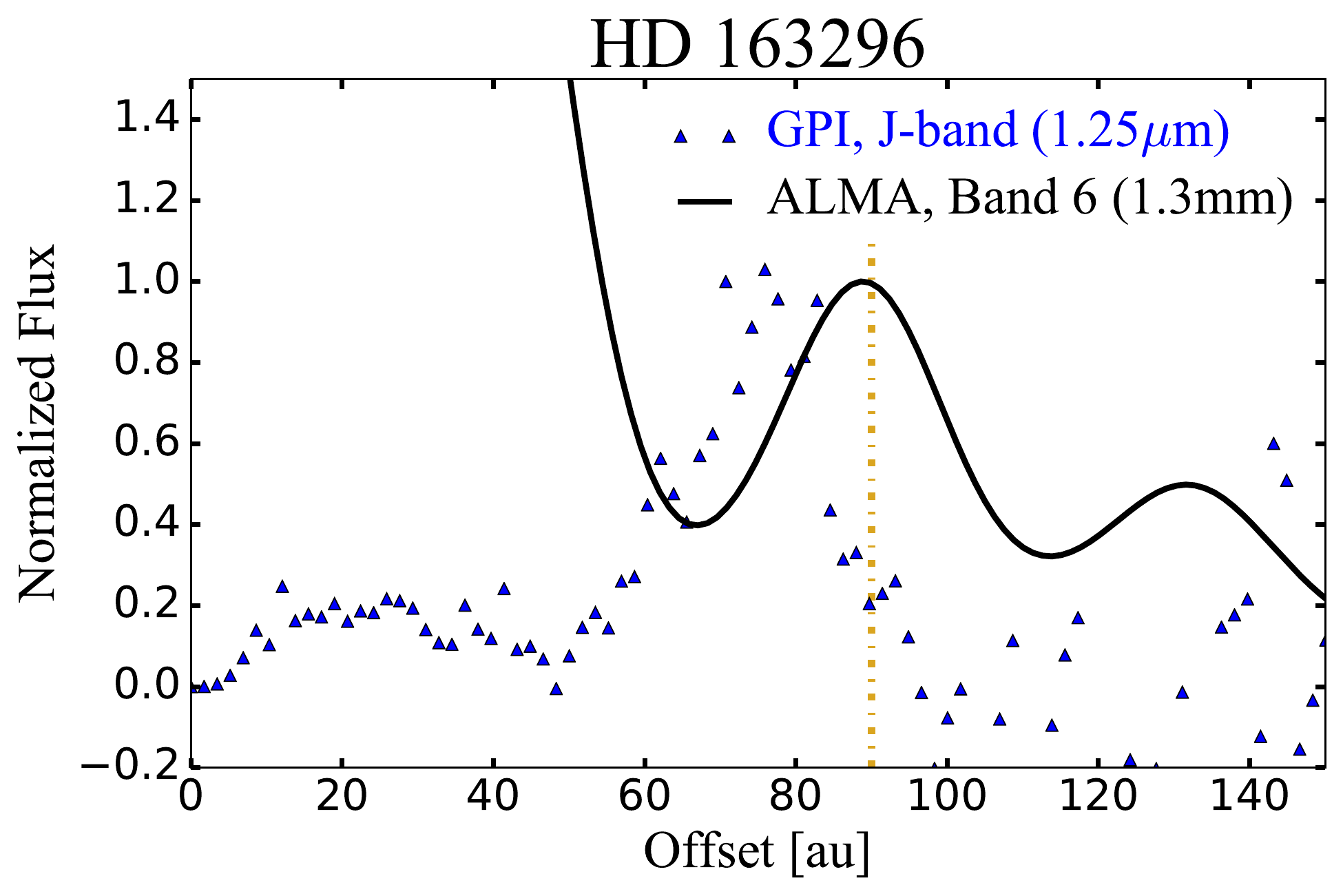}
   \end{tabular}
   \caption{Left panel: comparison of the radial intensity profile of TW\,Hya from SPHERE \citep{boekel2017} at 1.65\,$\mu$m vs. the submillimeter emission at 870\,$\mu$m from ALMA \citep{andrews2016}. Both profiles were obtained by azimuthally averaging the surface brightness intensity, and we normalize it to one of the peaks of emission (peak at $\sim$0.\arcsec8 for the SPHERE profile and peak at $\sim$0.\arcsec5 for the ALMA profile). The profile from SPHERE is multiplied by $r^2$ to compensate for the stellar illumination. The vertical line represents the location of the CO ice line at 20\,au \citep[assuming a distance of 59.5\,pc;][]{gaia2016}. Right panel: comparison of the radial intensity profile of HD\,163296 from GPI at 1.25\,$\mu$m \citep{monnier2017} vs. the millimeter emission at 1.3\,mm from ALMA \citep{isella2016}. Both profiles were obtained taking a cut at the position angle of the disk ($\sim135^\circ$), and we normalize it to one of the peaks of emission (peak at $\sim$0.\arcsec6 for the GPI profile and peak at $\sim$0.\arcsec7 for the ALMA profile). The profile from GPI is multiplied by $r^2$ to compensate for the stellar illumination. The vertical line represents the location of the CO ice line at 90\,au \citep[assuming a distance of 122\,pc;][]{ancker1997}.}
   \label{comparison_TWHya_HD163}
\end{figure*}

\subsection{Differentiating between Other Gap-opening Mechanisms}

There are several possibilities for the origin of rings and gaps in protoplanetary disks, including radial variations of the disk viscosity and planet-disk interaction. However, different observational diagnostics can give insights to distinguish between these scenarios.

Models of planet-disk interaction predict that when the planet is massive enough to open a gap in the gas surface density, there is trapping of millimeter-sized particles at the outer edge of the gap, where the density and hence the pressure increases outwards \citep{rice2006, zhu2011, pinilla2012}. Depending on the planet mass (which changes the pressure gradient) and disk viscosity, there is a critical grain size that can be trapped. In general, the micron-sized particles are not effectively trapped, and they can be distributed throughout the disk showing a smooth distribution or a shallower and smaller gap at short wavelengths than at millimeter emission \citep{ovelar2013, dong2015}. This is an important difference between our current models of ice lines, where a shallower gap is obtained at the millimeter emission. For instance, there is a dip of emission near the water ice line that is much more depleted at $1.6\mu$m than at 0.87 or 3.0\,mm. In addition, in the case of ice lines, the width of the gap or the separation between the rings is smaller at longer wavelengths.

For the gas distribution, in the case of massive planets, a deep gap is also expected in the \emph{total} gas surface density, for which depth and width again depend on the planet mass disk viscosity. In the case of ice lines, it is not expected that the \emph{total} gas surface density has  strong variations \citep[e.g.][]{ciesla2006, stammler2017}. The CO molecular line and its isotopologues  are usually used to infer the gas surface density distribution, but these can be strongly affected by the CO ice line \citep[e.g.][]{schoonenberg2017}.

In the case where the planet is not massive enough to open a gap in the gas surface density, the dust can still be shaped in  rings and gaps, due to changes in the gas velocities near the planet position, which can slow down the radial dust motion \citep{paardekooper2004, paardekooper2006, dipierro2016, rosotti2016}. As in the case of ice lines, the total gas surface density should not show any strong depletions. Nevertheless, the gaps in the case of low-mass planets are expected to be shallower and narrower for smaller grains than for large grains. As a consequence, measuring the depth and the width of an observed gap at near-infrared emission and millimeter emission is one of the keys to discerning between ice lines and low-mass planetary origin.

Alternatives for gap-opening processes are variations of the disk viscosity that can lead to different transport of angular momentum, creating bumps in the gas surface density that can lead to trapping of particles. The trapping mechanism is more effective for large millimeter grains, while the small grains are expected to be diffused and be smoothly disturbed in the disk  \citep{pinilla2012a}, opposite to our current results for the ice lines.

\subsection{Comparison with Current Observations and Future Perspectives} \label{sect:comp_obs}

\paragraph{TW\,Hya and HD\,163296.}
We discuss our results in the context of current observations of these two disks because they are very well studied in the literature, where the CO ice line has been claimed to be imaged by ALMA.  In addition,  there are constraints in the level of turbulence in these disks, making them excellent laboratories for discussing our current results.

In TW\,Hya, the ionization environment was investigated by \cite{cleeves2015}, suggesting a dead zone extending up to a distance of 60\,au. The CO ice line was suggested to be at 30\,au from observations of N$_2$H$^+$, which might be a chemical tracer of CO ice \citep[e.g.][]{qi2013}. However, \cite{hoff2017} demonstrated that the amount of CO in the gas phase can affect the abundance of N$_2$H$^+$, such that the  N$_2$H$^+$ column density peaks further outside the CO snow line. Assuming disk parameters of the TW\,Hya disk, \cite{hoff2017} found that the CO ice line should be located at 20\,au, in agreement with recent ALMA observations \citep{zhang2017}.

Near a distance of  20\,au there is a distinct gap at scattered light follow by a ring-like emission in the TW\,Hya disk \citep[see Fig.~\ref{comparison_TWHya_HD163};][]{rapson2015, boekel2017}, and ALMA observations also reveal a gap at 0.87\,mm that is narrower than the one observed at scattered light \citep[see Fig.~\ref{comparison_TWHya_HD163},][]{andrews2016}. This gap at 20\,au follows the trend found for the gap shape of our models when the disk viscosity is low \citep[$\alpha=10^{-3}-10^{-4}$, see also Fig.10 in][]{zhang2017}, in agreement with the possibility that the CO ice line is inside an MRI-dead region. It is  important to notice that our models are for a disk around a Herbig star and the CO ice line is much farther out (close to the outer edge) than in TW\,Hya. However, the observed shapes of the gap near 20\,au are similar to the results of our models at $\sim$8\,au, where the fragmentation velocity increases by one order of magnitude. The observed gap, in particular in the millimeter emission, is narrower than our prediction. In our models, the width is determined by where the fragmentation velocity changes and this determines whether the maximum grain size is dictated by fragmentation or drift. Our models may be neglecting variations of the fragmentation velocity due to, for instance, other main volatiles that change the stickiness efficiency near the CO ice line that can make the gap smaller.  

Additional observational insights can be obtained if we can infer the \emph{total} gas surface density. If a deep gap is also detected in the \emph{total} gas surface density at this location, this would suggest that the CO ice line is not responsible of the observed gap.

In HD\,163296, the level of turbulence has been constrained to be low \citep{flaherty2015}, and the CO ice line is located at $\sim90$\,au \citep{qi2015}.  \cite{guidi2016} suggested a gap followed by a ring feature in the millimeter emission just beyond the CO ice line, where there is also an excess of emission at polarized scattered light \citep{garufi2014, monnier2017}. The most recent ALMA observations from \cite{isella2016} show a strong ring-like signature near $\sim$90\,au as well (Fig.~\ref{comparison_TWHya_HD163}). The corresponding gap is deeper at scattered-light emission than at millimeter emission. In this case, the trend of a shallower and narrower gap at millimeter emission than at scattered- light emission is also observed (Fig.~\ref{comparison_TWHya_HD163}), following our current predictions for the ice lines.

Because these comparisons were done with polarized light intensity, we run as a test a 3D single scattering model for one of the cases (model~III with $\alpha=10^{-3}$), keeping the same disk parameters to calculate the polarized intensity profile at 1.6\,$\mu$m. The resulting shape of the radial profile of the polarized intensity is similar to the profile of the total intensity.

Both TW\,Hya and HD\,163296 have multiple rings and gaps, and each might have a different origin. Current observations support that the structures observed near the CO ice line are in good agreement with variations of the dust fragmentation efficiency. The other rings and gaps can originate, for instance, from embedded planets or viscosity variations.

\paragraph{Future observational perspectives.}
Detecting the water ice line directly from water lines is challenging because these lines are weak and they are at high energy levels \citep{notsu2016, notsu2017}. An indirect method to detect ice lines is to observe the effect that these locations have on the grain growth \citep[as shown in][]{birnstiel2010}, which can lead to strong variations of the emission at millimeter wavelengths and possible variations of the millimeter spectral index \citep{banzatti2015}.

\cite{cieza2016} observed with ALMA the protoplanetary disk around V883 Ori, which shows a break in the  intensity profile and an optical depth discontinuity at $\sim$42\,au, and they suggested that this break originates owing to the water ice line changing the dust properties and evolution. Because this protostar is experiencing an outburst in luminosity, this increases the disk accretion and temperature, pushing the location of the water ice line farther out than in a typical disk around a T Tauri or Herbig star. Future observations of the intensity at different wavelengths, covering different spectral types, are required to test the robustness of this theoretical prediction.

Observations that provide information about the \emph{total} gas surface density are complementary because near the ice lines strong depletions of the gas are not expected as in the case of massive planets or variations of the disk viscosity (e.g. at the outer edge of a dead zone). Moreover, the combination of optically thick emission that gives constraints on the disk temperate, together with observations at scattered-light and optically thin (sub)millimeter wavelengths, is an additional and crucial key to recognizing ice lines as potential origins of an observed ring or gap at a given location.

\section{Conclusions} \label{sect:conclusions}
In this paper we investigate the effect that different ice lines have on the dust density distribution at million-year time scales, by including radial variations of the fragmentation velocity at the ice lines of H$_2$O, CO$_2$ (or CO), and NH$_3$. Our findings are as follows:

\begin{enumerate}
\item Variations of the fragmentation properties of dust particles near ice lines can lead to visible gaps and rings, in particular at short wavelengths.

\item The amount of gaps depends on the number of ice lines considered. When only the water ice line is assumed, there is a break of the emission near the ice line at different wavelengths and independent of disk viscosity. The depth of the break or dip is higher for lower viscosity.

\item When the  CO$_2$ (or CO) and NH$_3$ are also assumed in the models, there is a clear gap between ice lines. The depth of the gaps depends on the disk viscosity, being deeper for lower viscosity. In addition, the formed gaps are narrower at longer wavelengths. The total number of rings and gaps can vary with wavelengths. For instance, in the case of three ice lines, there are two clear gaps and three bright regions in the emission at 1.6\,$\mu$m, but only one gap at the (sub)millimeter emission.

\item The general trend of our results is that gaps at the (sub)millimeter emission are shallower and narrower than at scattered light. Comparing the scattered-light and millimeter observations of TW\,Hya and HD\,163296, where the CO ice line has been observed, there are structures near the CO ice line (at 20 and 90\,au for TW\,Hya and HD\,163296, respectively) that are in agreement with our findings. This is opposite to the results expected by models of dust trapping by a giant planet embedded in the disk or when trapping is triggered by changes of disk viscosity, where a deeper and wider gap is expected at the (sub)millimeter emission.

\item In our models the gap width corresponds to the separation between ice lines, and therefore the gaps are very wide compared to the rings observed in disks around, e.g., TW\,Hya and HL\,Tau. However, any other volatile or mechanism that contributes to change the dust fragmentation velocities can lead to closer gaps and rings, or they can have a different origin.

\item In these models, we do not expect a strong change of the \emph{total} gas surface density near the ice lines. Massive planets or dead zones can lead to strong variations (gaps or bumps) in the gas surface density.

\end{enumerate}

\acknowledgements{We are very thankful to T.~Bergin and A.~Youdin for the enthusiastic discussions about this project. We thank S. Andrews, A. Isella, J. Monnier, and R. van Boekel for providing or making available the ALMA, GPI, and SPHERE data. P.P. acknowledges support by NASA through Hubble Fellowship grant HST-HF2-51380.001-A awarded by the Space Telescope Science Institute, which is operated by the Association of Universities for Research in Astronomy, Inc., for NASA, under contract NAS 5- 26555. T.B. acknowledges funding from the European Research Council (ERC) under the European Union's Horizon 2020 research and innovation program under grant agreement no. 714769. An allocation of computer time from the UA Research Computing High Performance Computing (HPC) at the University of Arizona is gratefully acknowledged.}

\begin{figure*}
 \centering
   	\includegraphics[width=18cm]{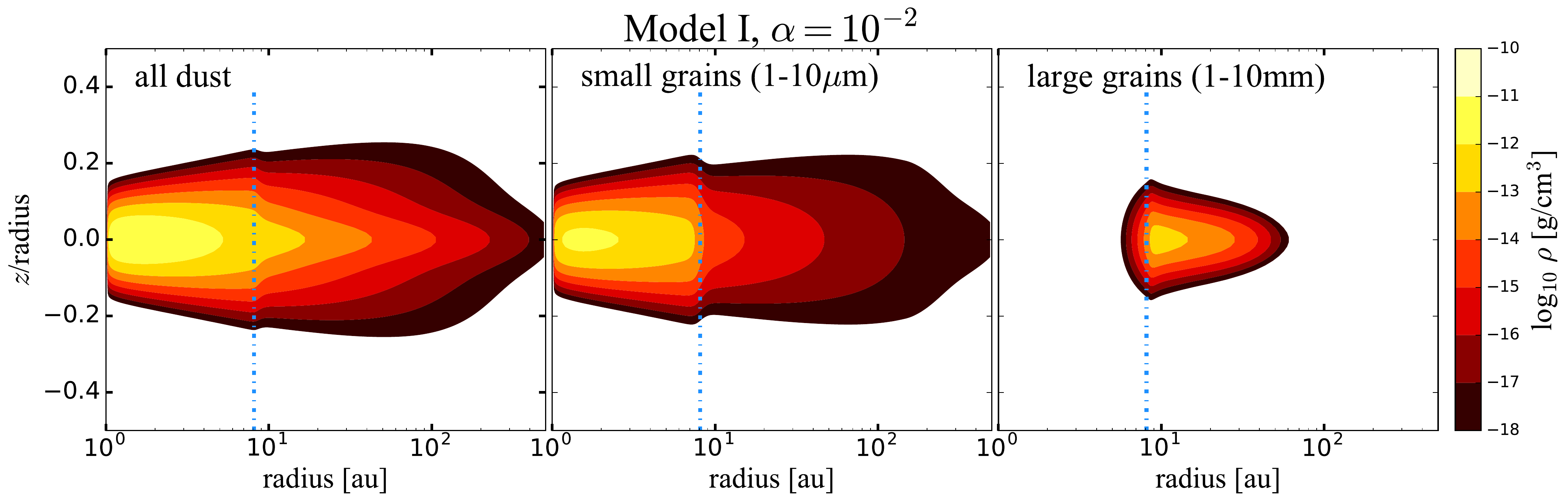}\\
	\includegraphics[width=18cm]{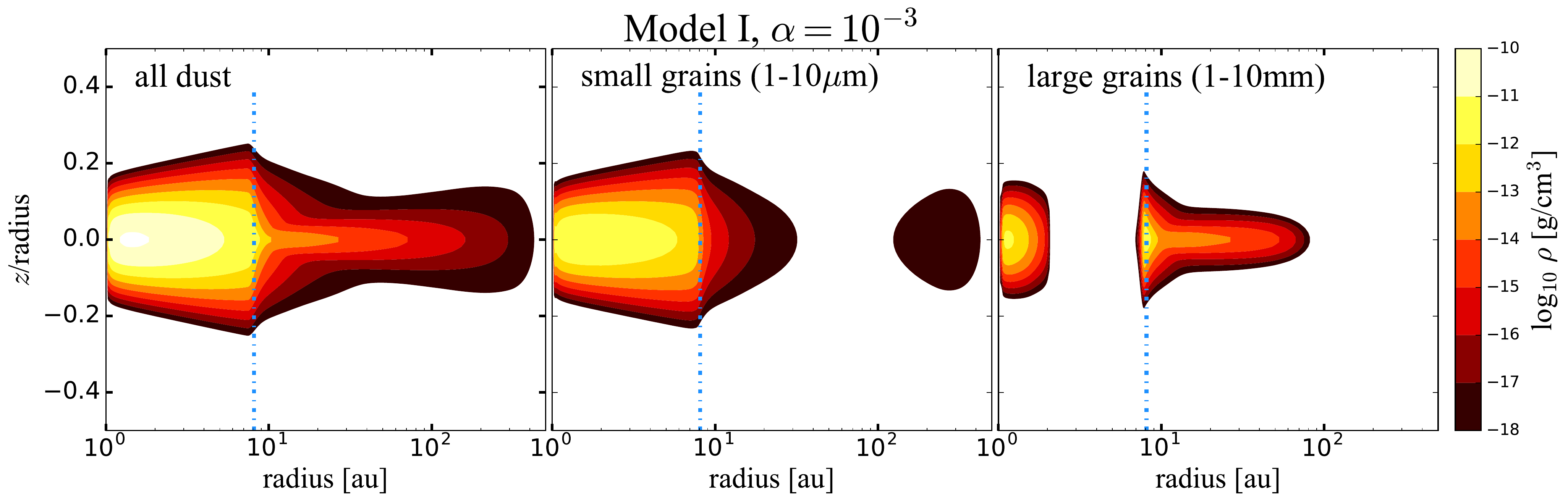}\\
	\includegraphics[width=18cm]{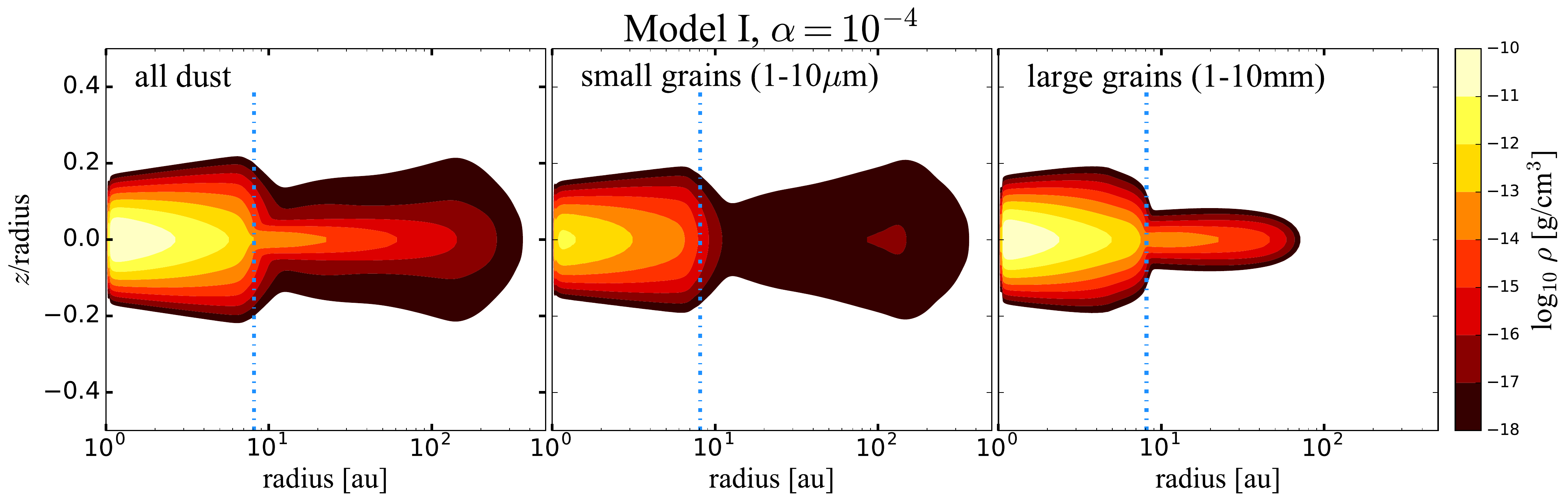}
    \caption{2D dust density distribution assumed for the radiative transfer calculations and obtained using Eq.~\ref{eq:volume_density}. We show the distribution for all the grain sizes (left panels), small grains ($a\in[1-10]\,\mu$m, middle panels), and large grains ($a\in[1-10]\,$mm, right panels).  All plots correspond to model~I (only H$_2$O ice line, vertical line) with different values of $\alpha=10^{-2}$ (top panels), $\alpha=10^{-2}$ (upper panels), and  $\alpha=10^{-4}$ (bottom panels).}
   \label{dust_2Ddensity_model1}
\end{figure*}

\begin{figure*}
 \centering
   	\includegraphics[width=18cm]{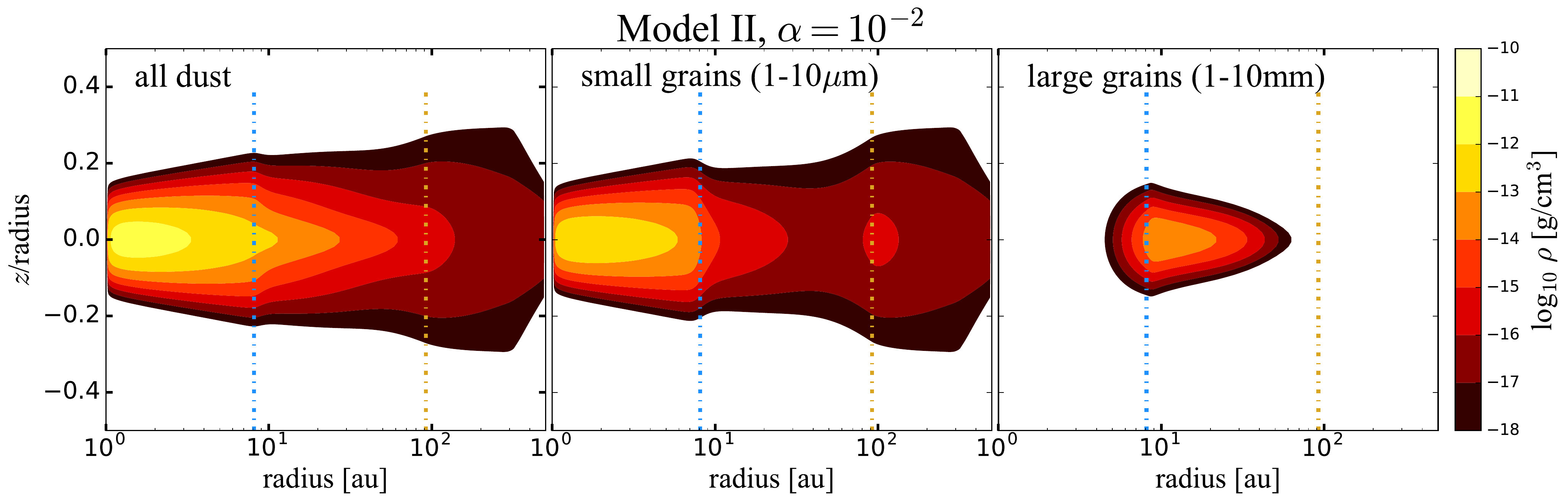}\\
	\includegraphics[width=18cm]{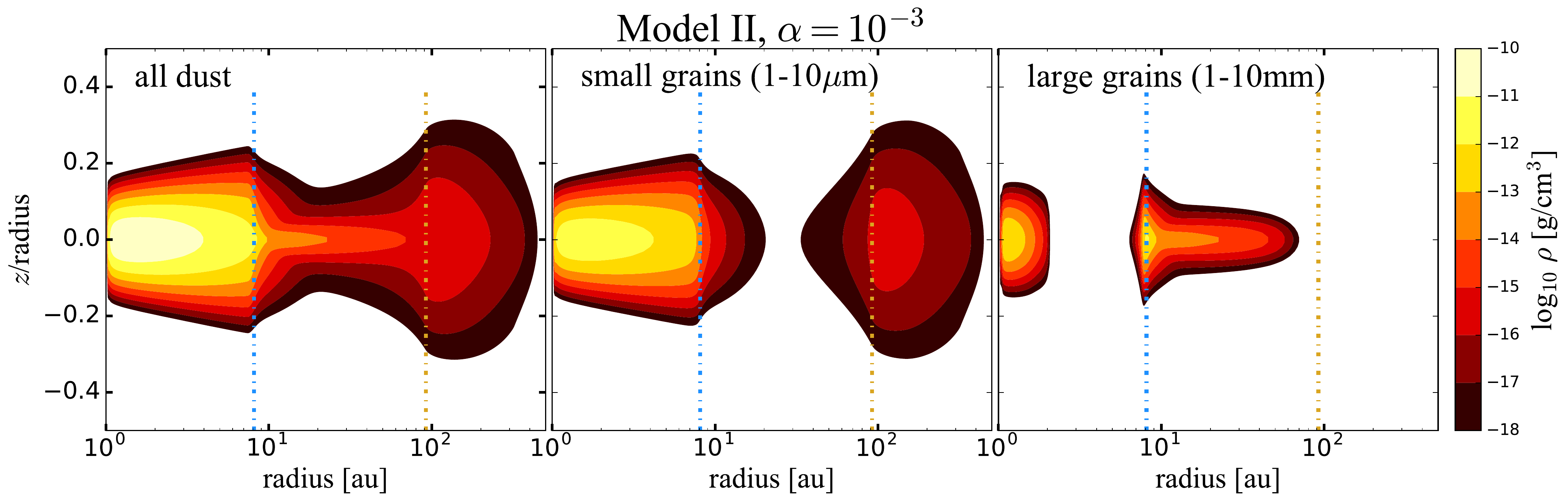}\\
	\includegraphics[width=18cm]{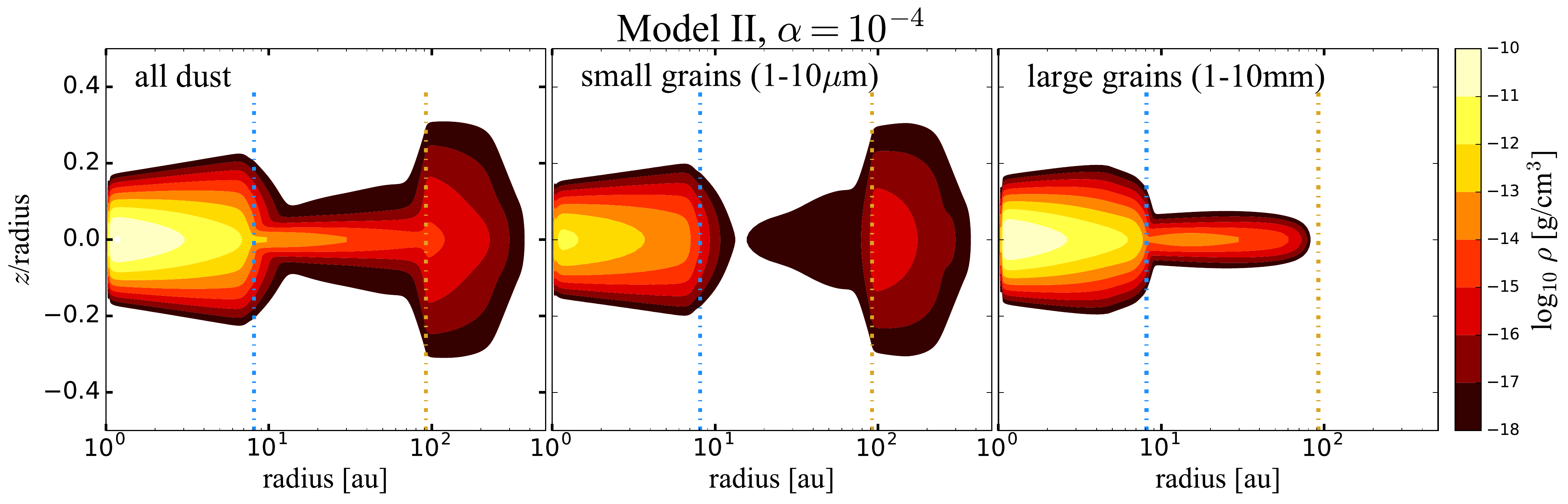}
    \caption{Same as Fig.~\ref{dust_2Ddensity_model1}, but for model~II (H$_2$O and CO$_2$ ice lines; vertical lines)}
   \label{dust_2Ddensity_model2}
\end{figure*}

\begin{figure*}
 \centering
   	\includegraphics[width=18cm]{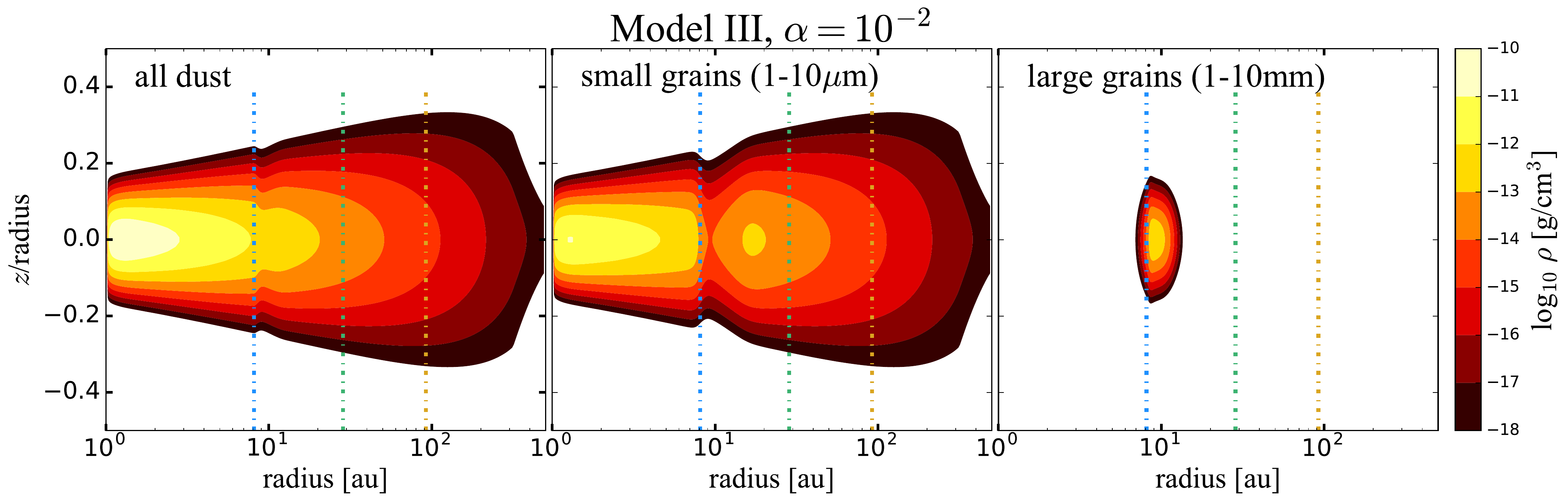}\\
	\includegraphics[width=18cm]{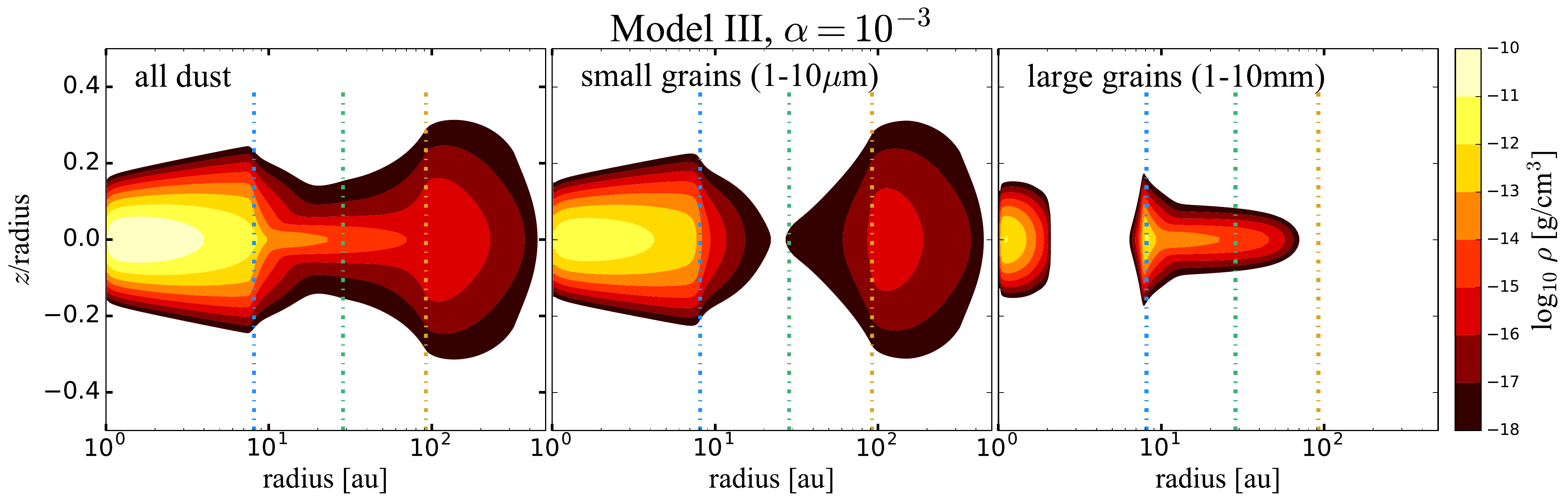}\\
	\includegraphics[width=18cm]{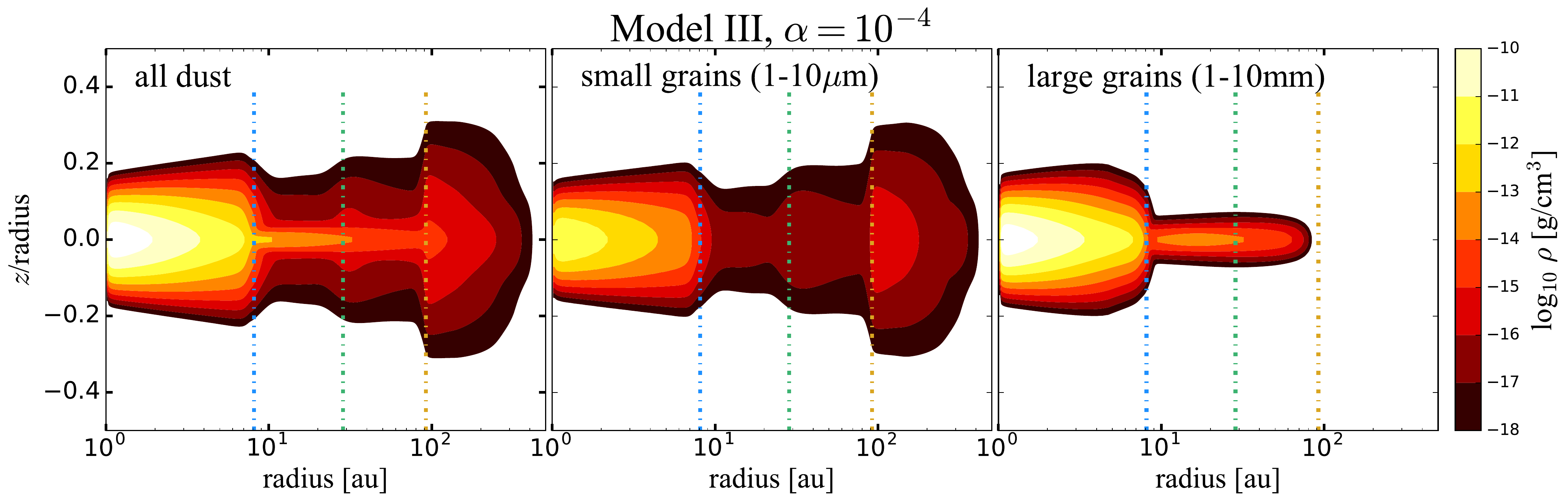}
    \caption{Same as Fig.~\ref{dust_2Ddensity_model1}, but for model~III (H$_2$O, CO$_2$, and NH$_3$ ice lines; vertical lines)}
   \label{dust_2Ddensity_model3}
\end{figure*}

\appendix{

Figures~\ref{dust_2Ddensity_model1}, \ref{dust_2Ddensity_model2} and \ref{dust_2Ddensity_model3} show the
2D $(r,z)$ dust density distribution assumed for the radiative transfer calculations and obtained using Eq.~\ref{eq:volume_density}, for small grains ($a\in[1-10]\,\mu$m), large grains ($a\in[1-10]\,$mm), and all grains, assuming different values of $\alpha$.
}

%%%%%%%%%%%%
%FIGURE 6
%%%%%%%%%%%%

\end{document}